\documentclass[american,11pt,3p,review,authoryear]{elsarticle}
\usepackage[T1]{fontenc}
\usepackage[latin9]{inputenc}
\usepackage{xcolor}
\usepackage{pifont}
\usepackage{bbding}
\usepackage{enumitem}
\usepackage{amsmath}
\usepackage{amssymb}
\usepackage{graphicx}
\usepackage{setspace}
\makeatletter

\providecommand{\tabularnewline}{\\}

\journal{International Journal of Forecasting}

\usepackage{amsmath,epsfig,amssymb,graphicx}
\usepackage{setspace,exscale,latexsym,srcltx}
\usepackage{longtable,pdflscape}
\usepackage[scanall]{psfrag}
\usepackage{rotating}
\usepackage[nottoc]{tocbibind}
\usepackage{dsfont}

\usepackage{tikz}
\usepackage{subcaption}

\usepackage{natbib}
\usepackage{breakurl} 

\usepackage{url}

\usepackage{rotating}
\usepackage{tocloft}
\usepackage{tikz}
\usepackage{color}
\usepackage{pifont}
\usepackage{tocbibind}
\usepackage{changepage}

\usepackage[colorlinks=true,linkcolor=blue,citecolor=blue]{hyperref}%
\usepackage[labelfont=bf]{caption}

\usepackage[english]{babel}
\usepackage{wrapfig}
\DeclareMathOperator*{\argmin}{arg\,min}

\usepackage{booktabs}

\usepackage{har2nat}

\usepackage{fancyhdr}	
\allowdisplaybreaks

\makeatother

\usepackage{babel}
\begin{document}

\begin{frontmatter}{}

\title{Conformal Prediction Interval Estimations with an Application to
Day-Ahead and Intraday Power Markets }

\author{Christopher Kath\textsuperscript{a,{*}}}

\ead{christopher.kath@rwe.com}

\author{Florian Ziel\textsuperscript{b}}
\begin{abstract}
We discuss the concept of Conformal Prediction (CP) in the context
of short-term electricity price forecasting. Therefore, we elaborate
the aspects that render Conformal Prediction worthwhile to know and
explain why this simple yet very effective idea has worked in other
fields of application and why its characteristics are promising for
short-term power applications as well. Its performance is compared
with different state-of-the-art electricity price forecasting models,
such as quantile regression averaging (QRA), in an empirical out-of-sample
study for three short-term electricity time series. We combine Conformal
Prediction with various underlying point forecast models to demonstrate
its versatility and behavior under changing conditions. Our findings
suggest that Conformal Prediction yields sharp and reliable prediction
intervals in short-term power markets. We further inspect the effect
each of the model components has and provide a path-based guideline
on how to find the best CP model for each market.
\end{abstract}

\address{\textsuperscript{a}University Duisburg-Essen, Chair for Energy Trading
and Finance }

\address{\textsuperscript{b}University Duisburg-Essen, House of Energy Markets
and Finance}

\ead{florian.ziel@uni-due.de}
\begin{keyword}
Energy forecasting, Prediction intervals, Electricity price forecasting,
Probability forecasting, Quantile regression, Linear models

\cortext[cor1]{Corresponding author}
\end{keyword}

\end{frontmatter}{}

\section{Introduction}

\label{sec:1}Our society is full of forecasts, whether it is for
economic data, the weather or customer demand. Unsurprisingly, this
general statement also applies to the energy industry. \citet{amjady2006energy}
describe the demand for accurate price predictions from two perspectives.
On the one hand, bilateral deals need to be realistically priced.
On the other hand, the necessity for reliable price estimations for
a) power producers to maximize their profit in power plant dispatch
and b) consumers to hedge and minimize their price uncertainty, becomes
evident. Thus, forecasting electricity prices is a highly active field
of research. Overviews on the status quo and available approaches
are supplied by \citet{aggarwal2009electricity,weron2014electricity}.
Whilst a variety of several point forecasts, i.e., the determination
of a concrete numerical estimate for the price, is already available
and being constantly improved by academics, uncertainty in forecasting
(e.g. expressed as prediction intervals) is only gaining more attention.
Inevitably, all forecasts involve uncertainty about their level of
preciseness, so why stop at the estimated price itself and not quantify
the unknown deviation that comes along with it? This is where prediction
intervals (PI) come into play. Based on the idea of an explicit consideration
of uncertainty, a prediction interval tries to identify a value range
that will most likely cover the observation. Unfortunately, extensive
studies of density or interval predictions are still relatively scarce.
The most prominent technique is quantile regression averaging (QRA)
in \citet{maciejowska2016hybrid,maciejowska2016probabilistic,nowotarski2014merging,nowotarski2015computing,uniejewski2018variance}.
It showed convincing results in various applications and marks the
current status quo for energy markets (more information on the mathematical
motivation is provided in chapter 4.3.2.). Other models are given
by bootstrapping (see a GARCH model in \citet{khosravi2013neural})
or quantile regression as in \citet{bunn2013analysis}. For a more
detailed discussion on probabilistic forecasting, the interested reader
might refer to a comprehensive study in \citet{nowotarski2017recent}.
\\
\hspace*{0.5cm}This paper contributes to this research field in the
following ways: We introduce a relatively unknown concept called Conformal
Prediction applied to day-ahead and intraday power prices. It is designated
to predict intervals based on errors, features weak assumptions on
data characteristics and is versatile with regards to the underlying
point prediction model. It might be seen as an expansion of an existing
point prediction estimator. But how does Conformal Prediction perform
under changing market conditions and in comparison to other approaches?
Can the approach deal with alternating point forecasts and the specialties
of hourly short-term prices? To find answers to these questions, the
remainder of this paper is structured as follows. To start with, we
thoroughly introduce the relatively new concept of Conformal Prediction
to the world of forecasting in section \color{blue} \ref{sec:2}\color{black}.
Before turning from a general Conformal Prediction toy example towards
a more dedicated electricity price scheme, we discuss the characteristics
of electricity prices based on three selected markets in section \color{blue} \ref{sec:3}\color{black}.
Once the theoretical foundation and time series description are dealt
with, we turn our attention towards the detailed models. We discuss
the general model setup in \color{blue} \ref{sec:4.1}\color{black},
our point forecasts in \color{blue} \ref{sec:4.2} \color{black}and
close the model description by elaborating our PI estimators in \color{blue} \ref{sec:4.3}\color{black}.
Section \color{blue} \ref{sec:5} \color{black} provides the results
of our empirical study based on several performance measures such
as the Winkler Score and pinball loss. In that context, we modify
a very basic model step by step until it equals Conformal Prediction
so that we can assess which specific aspect has the highest impact
on performance. Finally, we conclude our findings in section \color{blue} \ref{sec:6} \color{black}
and critically assess potential improvements for further research.

\section{The concept of Conformal Prediction}

\label{sec:2}Conformal Prediction (CP)\footnote{If we think in a broader sense, Conformal Prediction describes an
entire framework with different sub-models. For reasons of clarity
we will denote our sub-models as 'Conformal Prediction' as well. Hence,
the framework and the model specific definition are used analogously
in this paper.} describes an entire framework and was thoroughly analyzed for the
first time in \citet{gammerman1998learning} and later in \citet{shafer2008tutorial}
and \citet{vovk2005algorithmic} for both regression and classification
problems. The interested reader might also check \citet{kowalczewski2019normalized}
for an application based on this paper. Conformal Prediction was initially
introduced in an online or transductive manner, such that different
data realizations are iteratively presented to the learning algorithm.
This is not only computationally costly but also less practice-oriented.
Many real-world applications require batch processing, meaning that
there is one learning set of historical observations and a function
that tries to derive a generic rule applicable to new data. Inductive
Conformal Prediction translates the transductive approach into a batch
or inductive setting. Please note that we refer to the batch case
for regression problems when mentioning CP. But what renders CP special
and why should forecasters know about it? We firstly address some
pros:
\begin{itemize}[label=\CheckmarkBold{}]
\item CP yields valid prediction intervals that meet the designated confidence
level $1-\alpha$. The user predefines the desired confidence level. 
\item Only the weak assumption of exchangeability is made, no assumption
of underlying distributions is required. 
\item CP is model-agnostic and can be coupled with every singular prediction
model as it solely uses the final prediction of a classification or
regression model.
\item The framework itself offers high versatility with its applications
in regression, classification or an online or batch setting. It post-processes
point or classification model estimates and is independent from the
underlying point forecast model characteristics and assumptions.
\item CP, by definition, computes prediction intervals in an out-of-sample
manner which reduces the risk of overfitting. To further decrease
the risk of overfitting, CP can be coupled with sampling techniques.
Compared to bootstrapping approaches, CP only needs sampled data once
which renders its computation very fast in comparison to usual bootstrapping
where many iterations are required.
\end{itemize}
Besides these useful properties, CP also has some drawbacks that need
to be kept in mind when thinking about a possible application:
\begin{itemize}[label=\ding{56}]
\item CP requires more historical observations than other models since
we fit a point prediction model, compute its out-of-sample forecasts
and derive the interval from it. As a rule of thumb, one needs 25\%
- 50\% more observations.
\item CP is not suitable for time-series with strong structural breaks.
We need to split the time-series for computing point predictions and
intervals. If there is a regime-switch for instance, the assumption
of exchangeability is not valid anymore and we might end up fitting
intervals based on an entirely new regime but still assuming the predictions
to be valid.
\item If the number of available historical observations is limited, CP
is heavily influenced by outliers or clustering of distribution tail
events. Therefore, forecasters must carefully inspect the time-series
before fitting any model.
\item CP computes symmetric prediction intervals whereas other approaches
such as quantile regression or empirical error distribution based
approaches separately focus on each quantile. While CP is superior
if the true error is approximately symmetrically distributed, it could
cause problems in an asymmetric scenario.
\end{itemize}
The most crucial aspect is the extension characteristic. Like an additional
layer, CP adds an interval estimate to an existing point forecasting
model. A core principle of this second layer is the existence of a
non-conformity score $\lambda_{i}$ with $i$ being an index for the
number of the observation after sampling. It determines how uncommon
an observation is in comparison with the real value. More information
on the index notations is provided by Figure 1. Suppose we have (according
to \citet{johansson2014regression})
\begin{itemize}
\item A dataset containing historical observations $\mathcal{Z}=\left\{ \left(\mathbf{x}_{1},y_{1}\right),\ldots,\left(\mathbf{x}_{L},y_{L}\right)\right\} $
that we randomly split into portions $\pi$ and $1-\pi$- in our case
- portions of 75\% and 25\%. The split ratio was determined in a limited
hyper-parameter tuning and worked best for our case:
\begin{enumerate}
\item A training set (portion $\pi$ of $\mathcal{Z}$)\\
$\mathcal{Z}_{\mathrm{train}}$$=\left\{ \left(\mathbf{x}_{1},y_{1}\right),\ldots,\left(\mathbf{x}_{M},y_{M}\right)\right\} $
\item A calibration set (portion $1-\pi$ of $\mathcal{Z}$)\\
$\mathcal{Z}_{\text{calib}}=\left\{ \left(\mathbf{x}_{M+1},y_{M+1}\right),\ldots,\left(\mathbf{x}_{L},y_{L}\right)\right\} .$
\end{enumerate}
\item A random forecast model that exploits $\mathcal{Z}_{\mathrm{train}}$
for training and yields estimate $\hat{y}_{L+1}$. Please note that
we train on $\mathcal{Z}_{\mathrm{train}}$ and utilize the data of
$\mathcal{Z}_{\mathrm{calib}}$ to obtain unbiased out-of-sample estimates
$\hat{y}_{M+1},\ldots,\hat{y}_{L}$.
\item The simplest non-conformity score $\lambda_{i}=\left|y_{i}-\hat{y}_{i}\right|$,
computed from the estimates $\hat{y}_{M+1},\ldots,$ $\hat{y}_{L}$,
in $\mathcal{Z}_{\mathrm{calib}}$. A forecaster could also apply
other non-conformity definitions tailor-made to the prediction problem
at hand, CP is not limited to the absolute error here.
\end{itemize}
\begin{figure}
\begin{centering}
\includegraphics[scale=0.5]{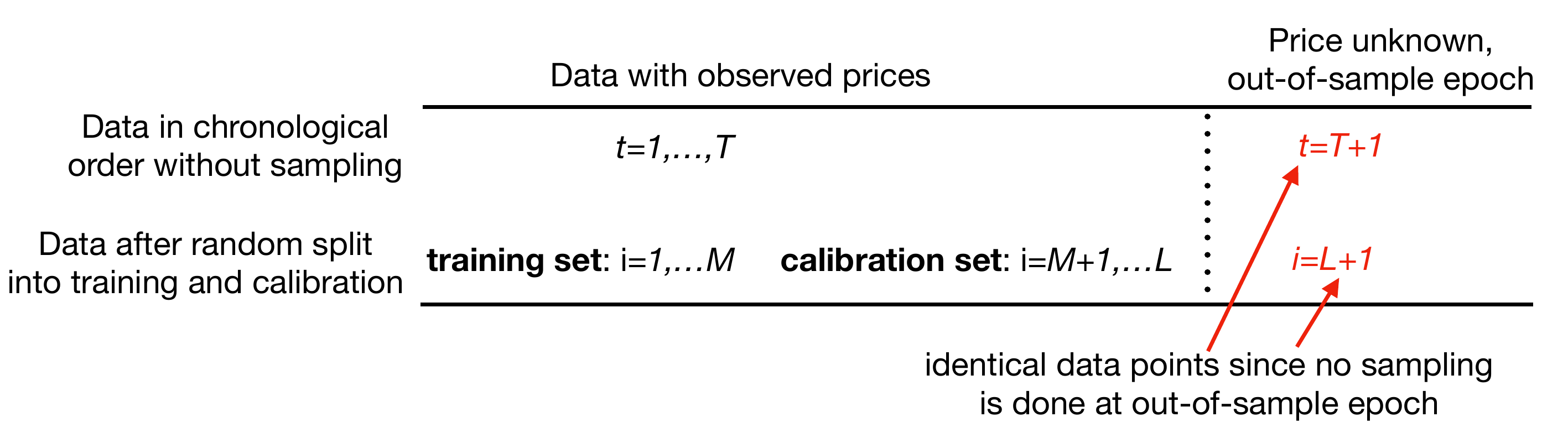}\caption{Detailed description of applied indices. Please note that there is
a different index notation for sampled and non-sampled data.}
\par\end{centering}
\end{figure}
\begin{figure}[t]
\centering{}\includegraphics[scale=0.4]{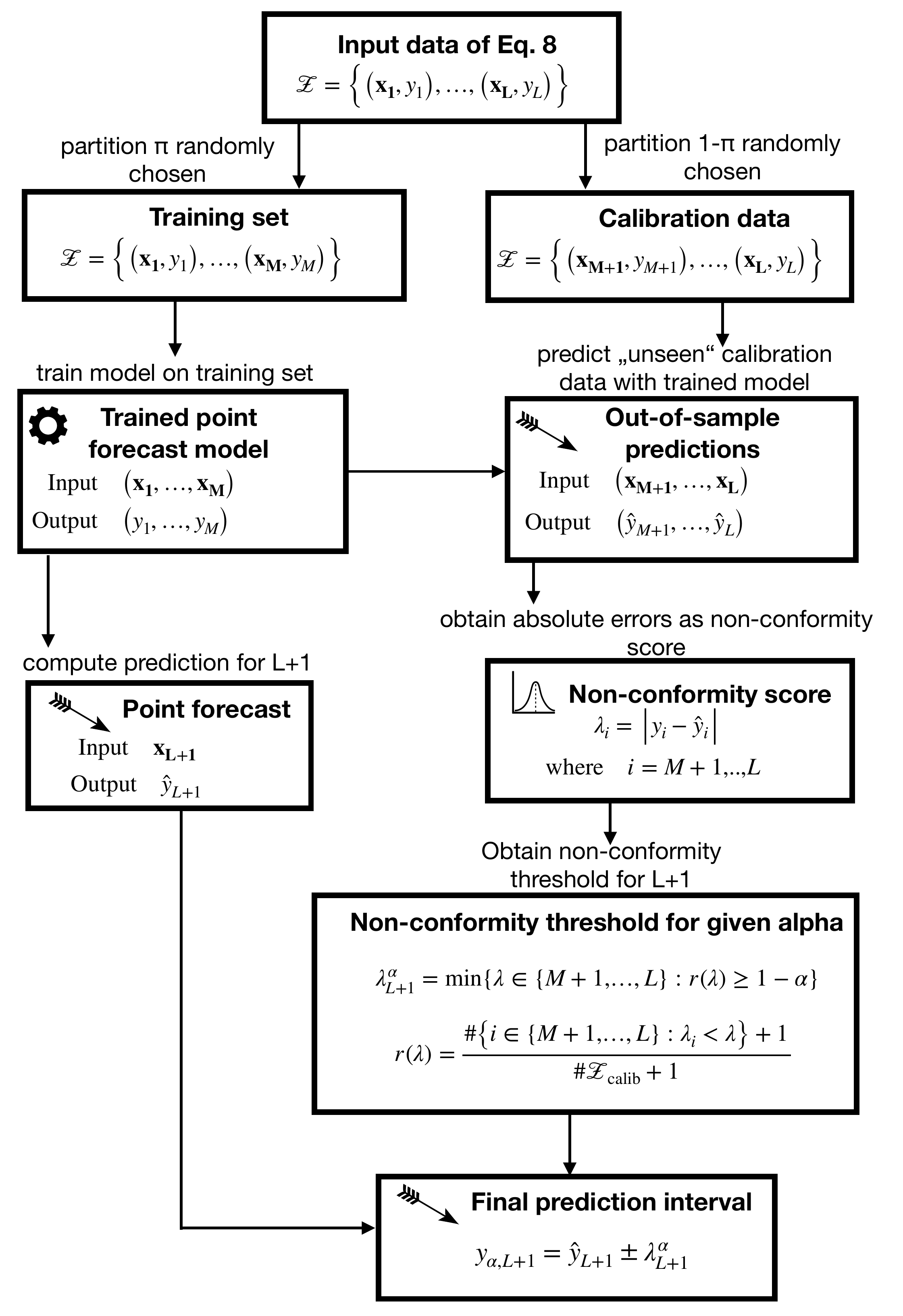}\caption{Schematic representation of Inductive Conformal Prediction. Detailed
information on input data per market is mention in Equation 6 as well
as section 3.1. The detailed hyper-parameters per model are described
in Appendix A. We use '\#' to denote cardinality. }
\end{figure}
The random split into training set $\mathcal{Z}_{\mathrm{train}}$
and calibration set $\mathcal{Z}_{\text{calib}}$\footnote{The notation of training and calibration is sometimes also used for
training and parameterization of models. We use the term 'calibration'
exclusively for Conformal Prediction and will use parameterization
whenever we want to express that a model needs to be tuned to identify
its optimal parameters.} is essential since we explicitly fit a model on $\mathcal{Z}_{\mathrm{train}}$
and exploit $\mathcal{Z}_{\mathrm{calib}}$ in an out-of-sample context.
It is important to mention that we only use $1,\ldots,M$ for training
the point predictors. This could lead to issues when the entire set
of historical observations is sparse. In the case of $M$ being very
small or seasonally influenced, CP might not be suitable anymore as
the underlying forecasting models are not trained properly. It is
hard to make a general recommendation as the necessity of data points
depends on the forecasting problem, but for electricity time series
we have found that a minimum of 1-2 months shall be available but
a year is even better to tackle all aspects of seasonality (more details
to be found in \citet{hubicka2018note,marcjasz2018selection}). If
one is uncertain about a proper choice of the hyper-parameter $\pi$,
we recommend to carry out a grid-search with, for instance, steps
of 10\% from 10/90 to 90/10 and evaluate which choice of portion yields
the best errors.\\
\hspace*{0.5cm}The point forecast model is trained to minimize the
error made with $\mathcal{Z}_{\mathrm{train}}$. Only considering
$\mathcal{Z}_{\mathrm{train}}$ results in a construction of intervals
on the basis of explicitly minimized in-sample errors and causes an
unrealistic estimation for unknown data. It does not reflect the model
behavior in an out-of-sample environment and could overfit the prediction
interval. Another comment must be made on our choice to split $\mathcal{Z}_{\mathrm{train}}$
and $\mathcal{Z}_{\text{calib}}$ in a random way. \citet{johansson2014regression}
only mention the partitions to be disjointed. However, under exchangeability,
the order of each element of $\mathcal{Z}$ must not matter. In case
of a seasonal time series with yearly, monthly or even weekly patterns,
the order makes a difference. Hence, a random split ensures the assumption
of exchangeability and prevents 'seasonal overfitting', i.e., training
only with data stemming from a specific time of the year. All in all,
random sampling is supposed to make the model more robust. A detailed
depiction of the interaction of both models, their input parameters
and the computation of intervals is provided by Figure 2.\\
\hspace*{0.5cm}Minding Figure 2, the main task of CP is to compute
a non-conformity score $\lambda_{L+1}^{\alpha}$ for the first out-of-sample
instance, i.e., the first epoch where an actual prediction is needed.
$\lambda_{L+1}^{\alpha}$ provides a probabilistic threshold so that
the non-conformity score for the true value $y_{L+1}$ will not exceed
$\lambda_{L+1}^{\alpha}$ with confidence $1-\alpha$. The threshold
value $\lambda_{L+1}^{\alpha}$ is identified by iterating through
all known $\lambda_{i}$ values and identify the smallest one under
a confidence level restriction in:
\begin{align}
{\lambda_{L+1}^{\alpha}=} & \min\{\lambda\geq0:r(\lambda)\geq1-\alpha\},\\
\mathrm{with}\,r(\lambda)= & \frac{\#\left\{ i\in\left\{ M+1,\ldots,L\right\} :\lambda_{i}<\lambda\right\} +1}{\#\mathcal{Z}_{\mathrm{calib}}+1},\nonumber 
\end{align}
where $\#\left\{ i\in\left\{ M+1,\ldots,L\right\} :\lambda_{i}<\lambda\right\} +1$
is the cardinality of all values $\lambda_{i}$ being smaller than
our chosen $\lambda$. In mathematical terms, $\lambda_{L+1,h}^{\alpha}$
is the smallest value that satisfies the condition $r(\lambda)\geq1-\alpha$.
A crucial aspect with that regard is the fact that $\lambda_{L+1}^{\alpha}$
automatically stems from the set of already computed non-conformity
scores over $M+1,...,L$. We only try to identify the smallest known
$\lambda$ value that satisfies the restrictions but do not compute
an entirely new numerical value, instead we set a threshold for future
values in $L+1$ based on past non-conformity scores. Also, note the
addition of ones in the counter and denominator. This simple trick
can be seen as a bias correction for small sizes of $M+1,\ldots,L$.
It does not make a difference with thousands of observations but a
simple case with a low cardinality as shown in Figure 3 shows the
effect. It allows for higher levels of $\alpha$ giving very limited
sample sizes.\\
 \hspace*{0.5cm}A toy example might be helpful in understanding this
concept. For the sake of simplicity, we assume 9 observations. The
instance $L+1$ is the one where we only face the given explanatory
variables $\mathbf{x_{\mathrm{\mathit{L+1}}}}$ and need to forecast
an interval for $y_{L+1}$. Figure 3 presents a solution minding Eq.
(1). 
\begin{figure}[t]
\centering{}\includegraphics[scale=0.52]{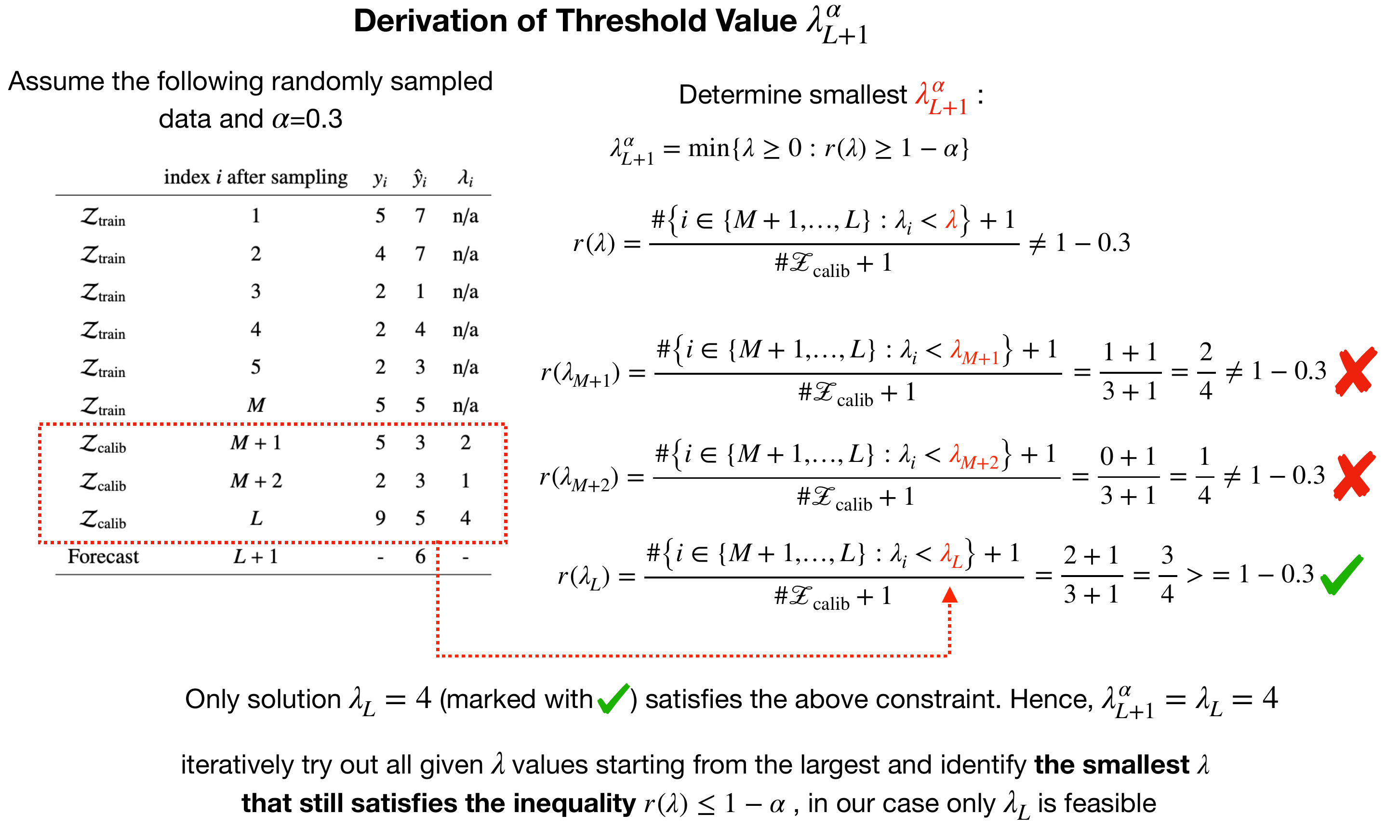}\caption{A toy example for the determination of Inductive Conformal Prediction's
threshold value that assumes $\alpha=0.3$ and a set of given forecasts
and observations. A solution for the threshold variable can be obtained
based on the formula in Eq. (1). More information on the depicted
indices $\ensuremath{1,\ldots,M,M+1,M+2,L,L+1}$ is provided in Figure
1.}
\end{figure}
 The result (in our toy example, only the largest value $\lambda_{L}=4$
meets the conditions) forms the interval around the point forecast
in 
\begin{equation}
\hat{y}_{L+1}\pm\lambda_{L+1}^{\alpha}.
\end{equation}
It is important to mention that Equation (2) presents the simple form
of Conformal Prediction, denoted as Inductive Conformal Prediction.
A more complex version with a normalized non-conformity score, called
Normalized Conformal Prediction, will be discussed in section 4.3.3.
This symmetric interval comprises the true price with confidence $1-\alpha$
under exchangeability in the underlying dataset. A less technical
explanation in \citet{shafer2008tutorial} exploits the law of large
numbers together with exchangeability. Suppose the exchangeable sample
space $\mathcal{Z}$ of size $N=M+L$ with subspaces $\mathcal{Z}_{n}=\{z_{1},\ldots,z_{n}\}$
for $n\leq N$ and the event $E_{n}=\left\{ y_{L+1}\text{\ensuremath{\notin}}[\hat{y}_{L+1}(\mathcal{Z}_{n})-\lambda_{L+1}^{\alpha},\hat{y}_{L+1}(\mathcal{Z}_{n})+\lambda_{L+1}^{\alpha}]\right\} $.
The event $E_{n}$ is $\alpha$-rare if the following holds true:
\begin{equation}
\mathbb{P}(E_{n}\mid\{z_{1},\ldots,z_{n}\})\leq\alpha.
\end{equation}
 We assume that the event $E_{n}$ given a random bag of data never
exceeds $\alpha$. \citet{shafer2008tutorial} also show that events
$E_{1},...,E_{N}$ are mutually independent which then implies that
$\mathbb{P}(E_{N-1}\mid E_{N})\leq\alpha$. Now if $N$ is large enough
and our event $E_{n}$ is an $\alpha-$rare event, the law of large
numbers proposes that events will only occur at portion $\alpha$
of $N$.\\
\hspace*{0.5cm}For a more technical derivation of such validity the
interested reader might study \citet{vovk2005algorithmic}. Whereas
other models fail to meet this requirement, CP leaves no concern about
validity but the width of the interval itself. It might yield broader
intervals than other approaches if the underlying point forecast model
is not precise or specific time-series characteristics are not regarded.
But why do we think that CP is suitable for electricity price forecasting?
Firstly, we discuss our time series in terms of scope and characteristics
and then present an adjusted CP scheme together with a set of point
forecasts and other PI expert learners.

\section{Data and case study framework}

\label{sec:3} We examine datasets that comprise electricity spot
prices from three different power markets: Nord Pool spot day-ahead
(year 2012 - 2013) prices, the German EPEX intraday market (year 2013-2016)
and the price track of the Global Energy Forecasting Competition 2014
(GEFCom 2014, data for years 2011 -2013). The choice of markets covers
geographical and chronological differences and provides insight into
the model performance under varying market conditions. At the same
time, we have chosen markets that are at least partially regarded
by other authors to have reproducible findings. Our outcome is comparable
to \citet{nowotarski2014merging} for the Nord Pool market. While
there is no EPEX intraday study available at the time of writing,
\citet{nowotarski2017recent} provide a benchmark for the GEFCom dataset.
The case study employs a common rolling estimation framework which
recalculates the model parameters on a daily basis and consequently
shifts the entire training, calibration and forecast window by 24
hours, as shown in Figure 4.
\begin{figure}
\centering{}\includegraphics[scale=0.35]{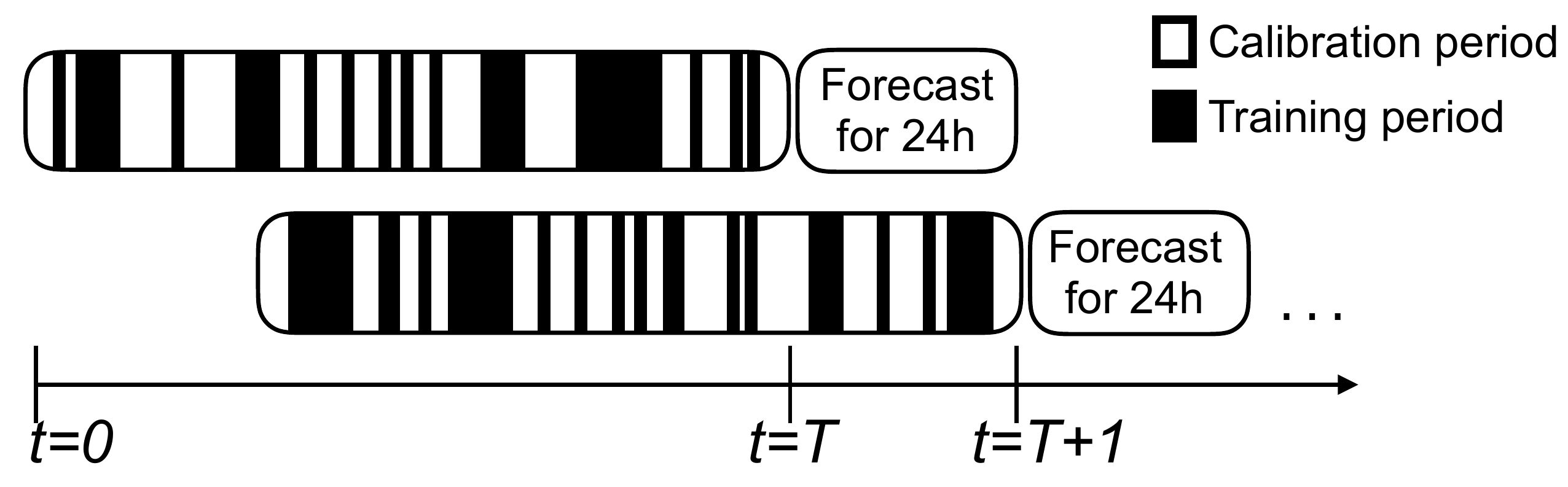}\caption{Out-of-sample rolling estimation scheme for our case study. The split
into training and calibration applies to NCP models only. Please also
note the random split depicted by changing areas of training and calibration.}
\end{figure}
The parameterization period (i.e., training and calibration phase)
spans 330 days and yields 182 days of Nord Pool forecasts and 831
daily intraday intervals respectively. GEFCom models yield 365 days
of out-of-sample predictions. We deliberately expand the estimation
window for intraday and GEFCom data to assess whether the models are
capable of reaching stable coverage ratios over a longer time horizon.
Conformal Prediction models and the naive benchmark are applied to
the entire parameterization period, while QRA is based on point forecasts
and cuts of eight weeks of parameterization data to train the quantile
regression model. For more information on reproducibility one might
check the data files mentioned under supplementary data. Also, note
that from now on we add index $h$ to reflect every single delivery
hour.

\subsection{Considered power markets}

The first time series we regard is the Nord Pool Spot system price
which is determined in a closed-form day-ahead auction at 12:00 CET.
It describes the unconstrained day-ahead price for the entire Nordic
bidding zone (e.g. Norway, Denmark, Sweden and Finland). It comprises
hourly spot electricity prices reported in EUR/MWh from 8.8.2012 -
31.12.2013. The price series can be obtained from the Nord Pool Spot
web page (\href{http://www.nordpoolspot.com}{http://www.nordpoolspot.com}).
Our case study refers to previous work of \citet{nowotarski2014merging}
which is why we replicate their basic setup: We calibrate the models
from 8.8.2012 - 3.7.2013 and report out-of-sample results for a 182-day
period spanning from 4.7.2013 - 31.12.2013. \\
\hspace*{0.5cm}German EPEX intraday trading prices reported in EUR/MWh
are the second short-term price series analyzed in this paper. While
the Nord Pool market allows entering a single round of bids establishing
the prices in a day-ahead auction, the EPEX intraday market is a continuous
one that is tradable up to 30 minutes\footnote{As of July 2017 EPEX allows to trade up to 30 minutes before delivery
from one German control area to another while the deadline for intra-control
area trades has shrunk to 5 minutes.} prior to delivery. Please note that this lead time changed as per
July 2015 from 45 to 30 minutes. We will consider the volume weighted
average price (VWAP) of all transactions for the specific delivery
hour. The data series can be obtained from the EEX historical data
service and ranges from 21.7.2013 - 30.9.2016. The initial training
and calibration window spans data from 21.7.2013 - 22.6.2014. We conducted
the out-of-sample test over 831 days to have valid findings not influenced
by any annual or seasonal effects. In contrast to the Nord Pool data,
we apply a set of external factors for the German intraday market.
The model is enriched with the ENTSO-E total load forecast obtainable
from \href{https://transparency.entsoe.eu/}{https://transparency.entsoe.eu/}
and estimated wind injection (freely available for download at \href{https://www.eex-transparency.com/}{https://www.eex-transparency.com/}).
These determinants are not only assumed to improve accuracy but also
increase complexity of the forecast model. Hence, we can validate
our model behavior under the usage of price information or multi-dimensional
regressor matrices. Please note that we have decided to ignore photovoltaics
production as this requires a more complex regression setup. Usually
one would leave a photovoltaics variable out of the model during night
times when there is no generation and add it in daylight hours. We
have sacrificed the additional input for the sake of a similar regression
setup in all three power markets.\\
\hspace*{0.5cm}
\begin{figure}[t]
\includegraphics[scale=0.56]{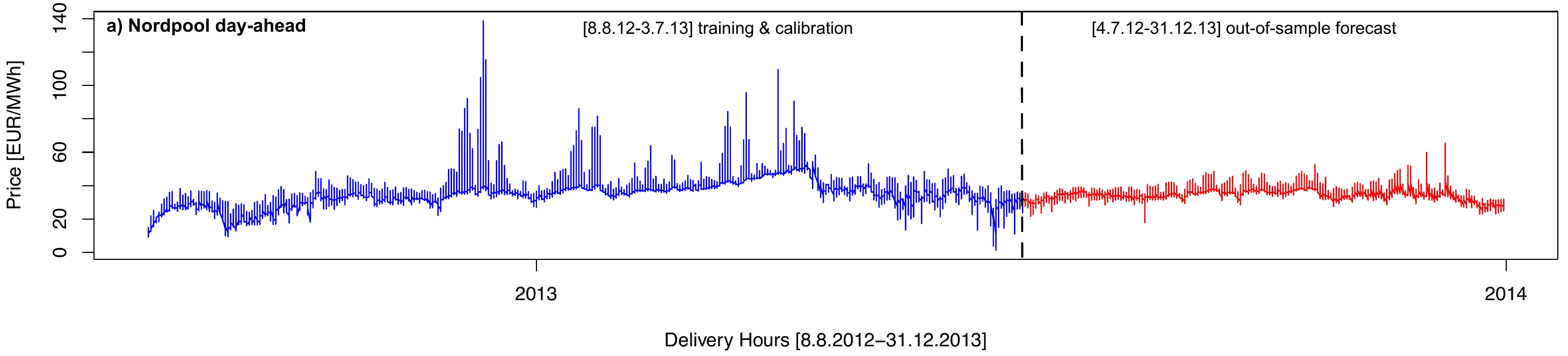}\\
\includegraphics[scale=0.56]{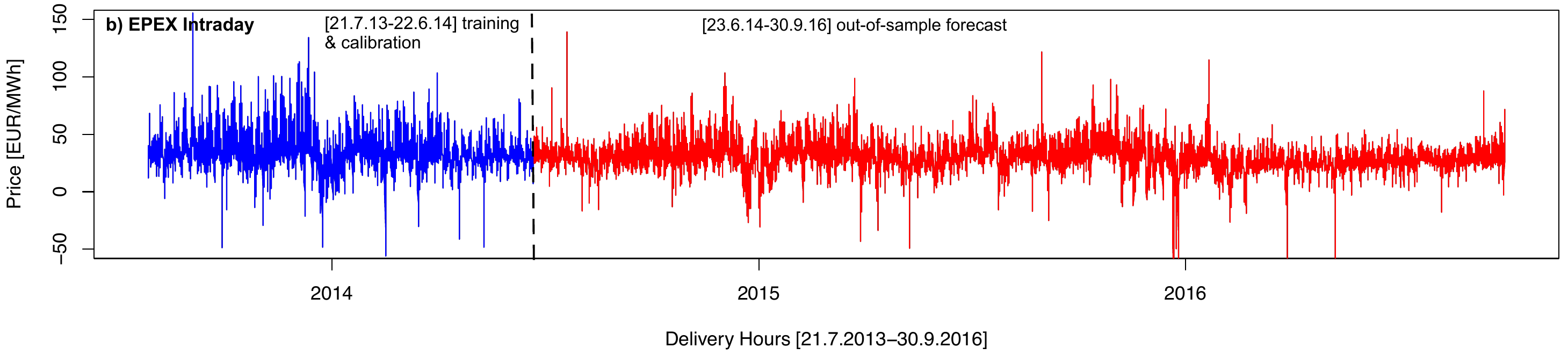}\\
\includegraphics[scale=0.56]{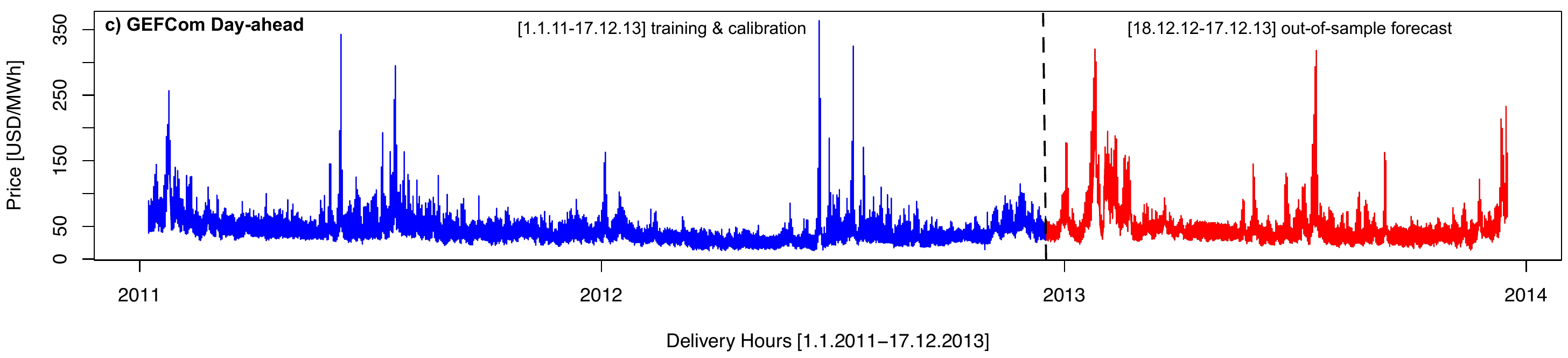}

\caption{Price plot of the Nord Pool, EPEX intraday VWAP and GEFCom day-ahead
time series separated into training and forecast sections. The blue
partition marks the initial training period that is consequently shifted
with each iteration of the rolling estimation. The red parts are used
for out-of-sample testing.}
\end{figure}
The last dataset stems from the Global Energy Forecasting competition
2014 and is available for download in the appendix of \citet{hong2016probabilistic}.
It covers hourly zonal prices in USD/MWh, zonal load forecasts and
system load predictions. The original market or exchange has never
been communicated by the authors but due to its usage in a large-scale
price forecasting competition it serves as a transparent, reproducible
benchmark dataset. We use all available data points which implies
a time period from 1.1.2011 - 17.12.2013. We follow the study in \citet{nowotarski2017recent}
and compute out-of-sample estimations from 18.12.12 - 17.12.13 to
have comparable findings.\\
\hspace*{0.5cm}But why have we chosen these price series? Figure
5 depicts how different the markets are. The three price series suggest
a mean-reverting tendency but the intraday (ID) time series features
higher volatility and negative prices. GEFCom data equals the intraday
data in volatility but shows an overall higher price level. 
\begin{table}[h]
\centering{}\centering{\footnotesize{}}%
\begin{tabular}{cccc}
 & {\footnotesize{}Nord Pool day-ahead} & {\footnotesize{}EPEX intraday} & {\footnotesize{}GEFCom 2014}\tabularnewline
\hline 
{\footnotesize{}mean} & {\footnotesize{}36.38} & {\footnotesize{}31.81} & {\footnotesize{}44.83}\tabularnewline
{\footnotesize{}SD} & {\footnotesize{}8.64} & {\footnotesize{}14.52} & {\footnotesize{}15.26}\tabularnewline
{\footnotesize{}1st quartile} & {\footnotesize{}32.55} & {\footnotesize{}24.04} & {\footnotesize{}33.42}\tabularnewline
{\footnotesize{}3rd quartile} & {\footnotesize{}39.90} & {\footnotesize{}38.98} & {\footnotesize{}53.93}\tabularnewline
{\footnotesize{}min} & {\footnotesize{}1.38} & {\footnotesize{}-155.52} & {\footnotesize{}12.52}\tabularnewline
{\footnotesize{}max} & {\footnotesize{}138.76} & {\footnotesize{}155.52} & {\footnotesize{}85.53}\tabularnewline
\hline 
\end{tabular}\caption{Descriptive statistics of the analyzed price series.}
\end{table}
 Table 1 supports the assumption of divergent characteristics. Both
the standard deviation (SD) and the interquartile range (IQR) are
much higher for intraday and GEFCom data. Interestingly, the spread
between the 1st and 3rd quantile is much higher with GEFCom prices,
while the difference between minimum and maximum is lower than the
other two markets. Hence, we do not only have entirely different price
series in terms of geographical and time characteristics, but also
the statistics support the impression of diversity.

\subsection{Pre-processing}

The applied time series exhibits hourly granularity which renders
a slight transformation necessary. Daylight saving time causes one
doubled hour and one missing value. We partly follow \citet{weron2007modeling}
and average the duplicate hours. The latter is computed using multiple
imputations as mentioned in \citet{buuren2011mice}. The multiple
imputations approach is also applied to all other missing data points
present in the time series. Figure 5 elicits concern for outliers
in our datasets. Conformal Prediction exploits descending errors and
could sacrifice preciseness to outliers. Hence, we tried the IQR based
Tukey method (see \citet{hoaglin2003john} for a more detailed description).
Outliers are defined by $1.5\textrm{*IQR}$ (like whiskers in common
box-plot graphics) and are replaced by multiple imputations after
removal. This process usually ensures greater generalization abilities
and less chance of wide intervals. It also marks an adjustment to
the underlying time series and needs to be treated very carefully.
Given our data and the parameterization window, forecast performance
(based on coverage, PI width and the Winkler Score) was decreased
by around 5 - 10 \%, which is why we decided to leave outliers unchanged
in our final time series. \\
\hspace*{0.5cm}Prices and input factors for regression are usually
transformed since many models demand stable variance. We apply a Box-Cox-based
power transformation denoted as Yeo-Johnson transformation. The great
benefit of that approach over the plain Box-Cox one is the capability
to deal with negative prices or zero values. It is defined in \citet{yeo2000new}
as
\begin{equation}
\psi(\eta_{h},y_{t,h})=\begin{cases}
((y_{t,h}+1)^{\eta_{h}}-1)/\eta_{h} & \mathrm{if\,\eta_{h}\neq0,y_{\mathit{t,h}}\geq0}\\
\mathrm{log}(y_{t,h}+1) & \mathrm{if}\,\eta_{h}=0,y_{t,h}\geq0\\
-((-y_{t,h}+1)^{2-\eta_{h}}-1)/(2-\eta_{h}) & \mathrm{if}\,\eta_{h}\neq2,y_{t,h}<0\\
\mathrm{-log}(-y_{t,h}+1) & \mathrm{if}\,\eta_{h}=2,y_{t,h}<0.
\end{cases}
\end{equation}
with an allowed range of $0\leq\eta_{h}\leq2$. Following \citet{yeo2000new},
the parameter $\eta_{h}$\footnote{Please note that we do not stick to the original notation in \citet{yeo2000new}
and denote the scaling parameter as $\eta_{h}$ as $\lambda_{i,h}$
is already used for Conformal Prediction.} is estimated by maximizing the maximum-likelihood function in 
\begin{align}
l_{n}(\theta_{h}|\mathbf{y_{h}}) & =-\frac{n}{2}\mathrm{log}(2\pi)-\frac{n}{2}log(\sigma_{h}^{2})-\frac{1}{2\sigma_{h}^{2}}\sum_{t=1}^{T}\left\{ \psi\left(\eta_{h},y_{t,h}\right)-\mu_{h}\right\} ^{2}\\
 & +(\eta_{h}-1)\sum_{t=1}^{T}\mathrm{sgn}(y_{t,h})\mathrm{log\left(\left|\mathit{y_{t,h}}\right|+1\right)},\nonumber 
\end{align}
where $\theta_{h}=\left(\eta_{h},\mu_{h},\sigma_{h}^{2}\right)$ and
$\mathbf{y_{h}}=(y_{1,h},...,y_{T,h})'$. The optimization in Eq.
(5) depends on the parameter $\eta_{h}$ and its impact on the population
mean $\mu_{h}$ and variance $\sigma_{h}^{2}$. The goal is to optimize
$\eta_{h}$ in such a way that a variance stabilizing transformation
is yielded. For more details on the optimization itself, the reader
can refer to \citet{yeo2000new}. Please note that we optimize each
$\eta_{h}$ individually per hour $h$, market (i.e., Nord Pool, GEFCom
and EPEX separately) and point forecast model (e.g. a different $\eta_{h}$
for each model described in section 4.2) using R's \texttt{caret}
package.

\section{Prediction models}

\subsection{General forecasting approach}

\label{sec:4.1}Our choice of input parameters is similar to those
in \citet{nowotarski2017recent} and \citet{nowotarski2014merging}
in order to have a comparable benchmark result. That being said, we
vary the input factors in our regression formula in Eq. (6) only slightly
to show the performance with (EPEX and GEFCom) and without (Nord Pool)
fundamental factors. The electricity price regression model itself
is given by Equation (6) with its connected separation of external
fundamental variables per market in Equation (7).
\begin{align}
y_{t,h} & =\beta_{1,h}+\underbrace{\beta_{2,h}y_{t-1,h}+\beta_{3,h}y_{t-2,h}+\beta_{4,h}y_{t-7,h}}_{\mathrm{\textrm{AR-terms}}}\\
 & +\underbrace{\beta_{5,h}y_{\min,t-1,h}+\beta_{6,h}y_{\textrm{max},t-1,h}}_{\mathrm{\textrm{non-linear\,effects}}}+\underbrace{\beta_{7,h}D_{Sat}+\beta_{8,h}D_{Sun}+\beta_{8,h}D_{Mon}}_{\mathrm{\textrm{daily\,dummies}}}\nonumber \\
 & +\underbrace{\beta_{10,h}\textrm{PC}\textrm{A}_{1,h}+\beta_{11,h}\textrm{PC}\textrm{A}_{2,h}+\beta_{12,h}\textrm{PC}\textrm{A}_{3,h}}_{\mathrm{\textrm{daily\,factors}}}\nonumber \\
 & +\underbrace{\beta_{13,h}y_{24,t-1,h}}_{\mathrm{\textrm{end-of-day\,effect}}}+\underbrace{\beta_{14,h}\delta_{t,h}}_{\mathrm{\textrm{threshold effect}}}+\underbrace{\beta_{n,h}\mathrm{\phi_{\mathit{t,h}}}}_{\mathrm{\textrm{fundamentals}}}+\varepsilon_{t,h},\nonumber 
\end{align}
with
\begin{equation}
\beta_{n,h}\mathrm{\phi_{\mathit{t,h}}:=}\begin{cases}
\mathrm{0} & \mathrm{for\,Nord\,Pool}\\
\underbrace{\beta_{15,h}\mathrm{\phi_{\mathit{1,t,h}}}}_{\mathrm{\textrm{zonal load}}}+\underbrace{\beta_{16,h}\mathrm{\phi_{\mathit{2,t,h}}}}_{\mathrm{\textrm{system load}}} & \mathrm{for\,GEFCom}\\
\mathrm{\underbrace{\beta_{15,h}\mathrm{\phi_{\mathit{1,t,h}}}}_{\mathrm{\textrm{load forecast}}}+\underbrace{\beta_{16,h}\mathrm{\phi_{\mathit{2,t,h}}}}_{\mathrm{wind\,forecast}}} & \mathrm{for\,EPEX},
\end{cases}
\end{equation}
where $y_{t-1,h}$, $y_{t-2,h}$, $y_{t-7,h}$ denote the prices of
the identical hour one, two and seven days ago, while $\beta_{n,h}$
is the respective regression coefficient. The indices $h$ and $t$
describe the hour and day of the underlying electricity price. Non-linear
price effects are considered by $y_{\textrm{min},t-1,h}$ and $y_{\textrm{max},t-1,h}$
being the minimum and maximum price of the previous day and $y_{24,t-1,h}$
the last known price, i.e., the price of hour 24 one day ago. The
terms $D_{Sat}$, $D_{Sun}$, $D_{Mon}$ are dummy variables (taking
a value of 1 in case of their occurrence) to capture the intra-week
term structure. $\textrm{PC}\textrm{A}_{k,h}$ is the $k-\mathrm{th}$
principal component of yesterday's 24 prices and comprises reduced
daily price information. A threshold variable $\delta_{t,h}$ picks
up the threshold model idea of \citet{nowotarski2014empirical} and
compares the mean of yesterday's daily prices with its equivalent
one week ago to determine low or high volatility price regimes. We
use the notation $\phi_{\mathit{t,h}}$ as a wildcard for all model-specific
fundamental inputs, i.e., none for Nord Pool, zonal and system load
forecasts for GEFCom and load and wind generation predictions for
EPEX intraday, as additionally described in Equation (7). Please note
that we intentionally sacrifice customization of the regression model
for the sake of harmonization and comparable findings. Without that
restriction one could also add day-ahead prices to the intraday regression
or discuss -depending on the time of forecasting- if early intraday
prices of the hour to be predicted would be a suitable addition.\\
\hspace*{0.5cm}The regression problem itself requires customization
of the underlying process; in our case the prediction of electricity
spot prices which inhibits certain specifics. Short-term electricity
time series feature manifold seasonality due to their hourly characteristics,
weekly effects and summer/winter times. We model each hour separately
as 24 individual processes to minimize hourly or base/peak effects.
The hourly granularity is reflected by the index $h$ in Eq. (6).
While this approach minimizes one source of heteroscedasticity, it
causes a different problem: Hourly interdependencies caused by ramping
costs or similar load events get lost. Traditional thermal power plants
exhibit boundaries like start-up times. These might cause one hour
to be profoundly affected by the preceding one. Many PI models ignore
this source of heteroscedasticity and disregard possible joint distributions
as mentioned by \citet{nowotarski2017recent}. We follow a different
approach. A principal component analysis (PCA) acknowledges these
effects in 
\begin{equation}
y_{t-1,h}\backsim\mathbf{\mathbf{\Lambda_{\mathrm{k,t}}}F_{\mathrm{k,t}}},
\end{equation}
where $\mathbf{\Lambda_{k,t}}$ are the load factors and $\mathbf{F_{k,t}}$
the principal components of yesterday's prices. The components comprise
all daily price information and are determined using all 24 hours.
Please note that $k=1,...,24$ because 24 hours yield 24 components.
As with conventional PCA, the first few factors comprise sufficient
information to be included. In our case, three components are utilized.
For another application of PCA in the context of electricity price
dimension reduction one can check \citet{raviv2015forecasting}.

\subsection{Individual point forecast models}

A common basis for many PI estimators are point forecasts in the form
of a simple regression where the actual price is a function of input
factors $\mathbf{x}_{t,h}$ and an error term $\varepsilon_{t,h}$.
We apply a variety of different models starting from a \textbf{Naive
}benchmark using past values as prediction, over an advanced penalized
linear regression model denoted as \textbf{Lasso,} to a K-nearest
neighbor (called\textbf{ KNN}) algorithm and a support vector machine
regression (\textbf{SVM}). All models are described in a more detailed
way in Appendix A.

\label{sec:4.2}

\subsection{Prediction interval models }

\label{sec:4.3}
\begin{figure}[!t]
\begin{centering}
\includegraphics[scale=0.55]{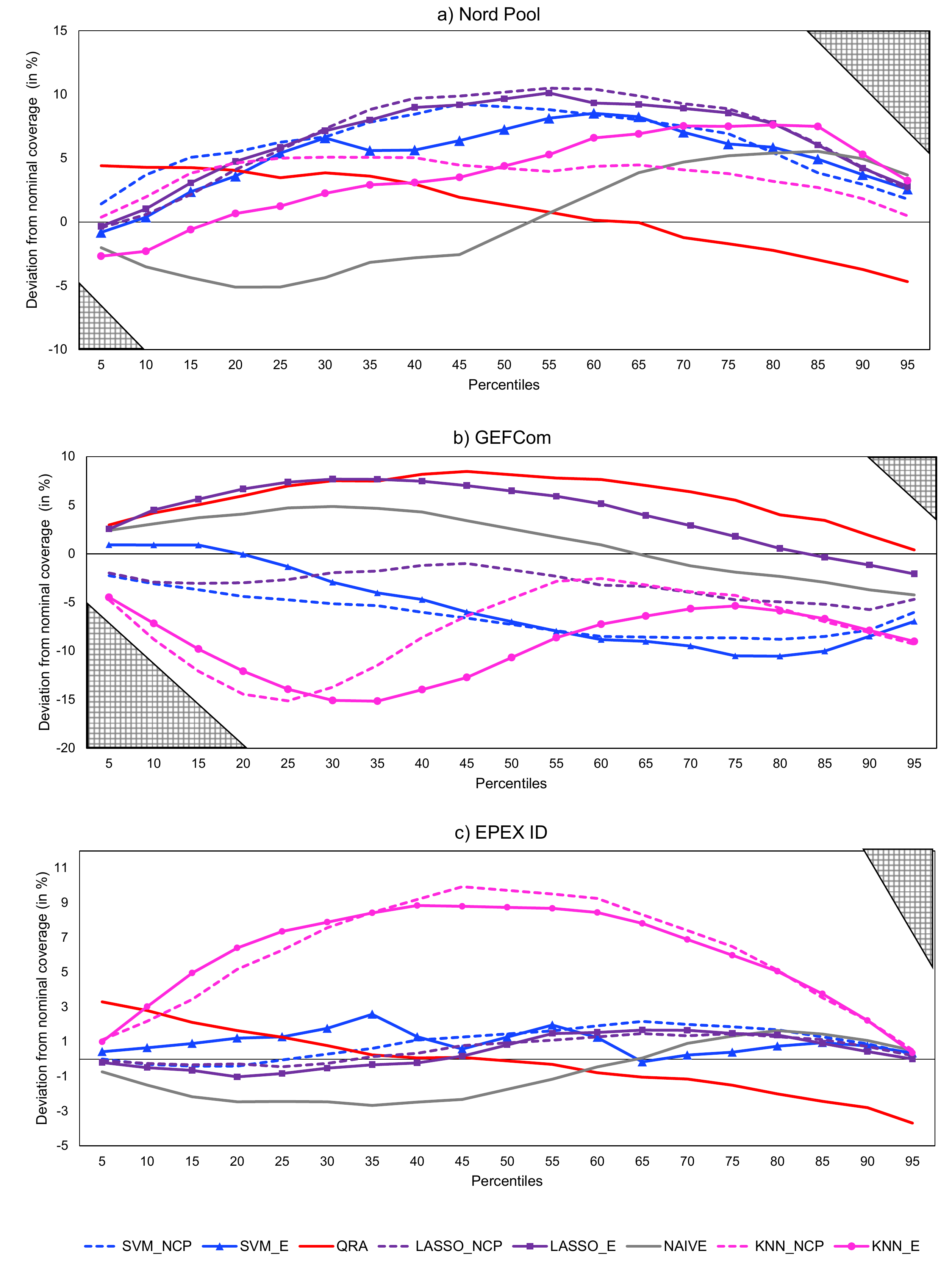}
\par\end{centering}
\caption{Differences between the empirical coverage of the prediction models
and the nominal coverage per percentile. While we have left out the
50th percentile in the initial calculation, it is depicted here using
interpolation. This ensures that we do not create any unwanted bias
through a steeper step between the 55th and 45th percentile. The hatched
gray area reflects the minimum possible coverage, i.e., the 5th percentile
cannot have a higher negative deviation than 5 percent.}
\end{figure}
A proper point forecast model is only the first step to retrieve prediction
intervals. A bit of attention must be paid to the intervals and its
notation. A $(1-\alpha)$ prediction interval implies that the interval
contains the true value with probability $(1-\alpha)$. Transferring
this idea to the calculation of quantiles leads to $\tau=\frac{\alpha}{2}$
for the lower and $\tau=(1-\frac{\alpha}{2})$ for the upper bound.
For instance, we calculate the 5 \% and 95 \% quantile which yields
a 90 \% prediction interval if the distance between the two quantiles
is regarded. A note must also be made on symmetry. Models can estimate
quantiles or PIs in a symmetric fashion by adding or subtracting from
a point forecast (see Eq. 13 for instance). Other models compute quantiles
independently such that we construct the PIs from two quantiles without
any point forecast in between.

\subsubsection{Empirical error distribution approach}

As a probabilistic benchmark, we introduce a simplistic, model-agnostic
approach called empirical error distribution (the suffix \textbf{\_E}
will be used in the following). Assume an expert learner (in our case,
Lasso, KNN or SVM or the naive model) with the described hyper-parameters
under section 4.2 and trained with the explanatory variables of Eq.
(6). We simply compute the forecast $\hat{y}_{t,h}$ for both the
calibration and training time window. Then, we calculate the forecast-individual
residuals $\varepsilon_{t,h}=\left|\hat{y}_{t,h}-y_{t,h}\right|$
(i.e., using the absolute error) and compute the sample quantile of
in-sample errors $\hat{q}{}_{\tau,h}$ over all $\varepsilon_{t,h}$
for $t=1,...,T$. Note that $q_{\tau,h}$ is not depending on time
$t$ but is calculated for the absolute error per hour $h$. We expand
the point forecast for the unknown data to $\hat{y}_{\tau,T+1,h}=\hat{y}_{T+1,h}\pm\hat{q}_{\tau,h}$
to retrieve the upper and lower bounds $\hat{y}_{\tau,T+1,h}$. This
procedure does not demand any assumptions on time series characteristics
nor require any greater effort and marks the minimum to be reached
for all other models. Please note that we do not use any sampling
for our quantile calculation such that one could argue that this automatically
leads to overfitting or the intervals being too narrow. This definitely
holds true for very small samples. However, given our sample size,
we follow the asymptotic theory and assume that we do not conduct
a large error. Besides, leaving out sampling - one of Conformal Prediction's
key factors - puts us in a position to specifically analyze its influence
in a dedicated study in sub-chapter 5.4. Another possible point of
criticism is the choice of the absolute error as the basis for the
quantile computation. We want to compute a symmetric estimator but
acknowledge that another residual definition for $\varepsilon_{t,h}$
could influence results, which is why we briefly touch upon asymmetric
quantiles in sub-chapter 5.4 as well. We assume the effect to be rather
minor as the residuals are nearly symmetric. In such a setting, there
is no substantial deviation if one calculates a quantile for absolute
values or their unadjusted equivalents.

\subsubsection{Quantile regression averaging}

Recent studies, as well as the GEFCom (see \citet{hong2016probabilistic}
for results), have shown how powerful the quantile regression averaging
(\textbf{QRA}) model of \citet{nowotarski2015computing} is. It stems
from the thought of combining forecasts to improve performance (e.g.
in \citet{bordignon2013combining,nowotarski2014empirical}). The approach
uses a set of individual point forecasts as an input for a quantile
regression. The output is a quantile of either forecast errors (see
\citet{maciejowska2016hybrid} for instance) or price levels (applied
in \citet{nowotarski2015computing}). The underlying problem formulation
is to be found in \citet{nowotarski2014merging} as
\begin{equation}
\ensuremath{q_{\tau,h}(y_{t,h})=\mathbf{\omega}_{\tau,h}'\hat{\mathbf{y}}_{t,h}+\varepsilon_{t,h},}\text{�}
\end{equation}
with $\hat{\mathbf{y}}_{t,h}$ being the vector of point forecasts
computed out of Eq. (6) by the different competing prediction models
mentioned in section 4.2. Hence, the vector comprises individual point
forecasts of the Lasso model, a support vector and a k-nearest neighbor
regression. Please note that the naive learner is not included in
the vector due to its simplistic character and the expected negative
effect on performance. The notation $\mathbf{\mathbf{\mathbf{\omega}}_{\mathrm{\mathit{\tau,h}}}}$
describes a vector of weights with which to multiply the model output.
The term $q_{\tau}(y_{t,h})$ denotes the conditional quantile of
the electricity price distribution given the user-specified nominal
coverage in $\tau$. The weights are determined by an optimization
in
\begin{gather}
\mathrm{\argmin_{\omega}\left[\sum{}_{t=1}^{L}\rho_{\tau}(\mathbf{y}_{\mathit{t,h}}-\omega'\hat{\mathbf{y}}_{\mathit{t,h}})\right],}
\end{gather}
where $1,...,\mathrm{L}$ describes the in-sample period and $\rho_{\tau}(z)=(\tau-\mathbb{I}_{\{z<0\}})z$.
Equation (10) is equivalent to a likelihood function of a linear regression
with asymmetric Laplace-errors and yields numerical values for the
upper and lower bounds. Please also note that QRA does not explicitly
account for heteroscedasticity. It necessitates point forecast estimates
as input factors but if these models do not consider the different
price realizations of weekdays and hours, the model might end up biased
for electricity prices.

\subsubsection{Normalized Conformal Prediction}

\begin{figure}
\begin{centering}
\includegraphics[scale=0.48]{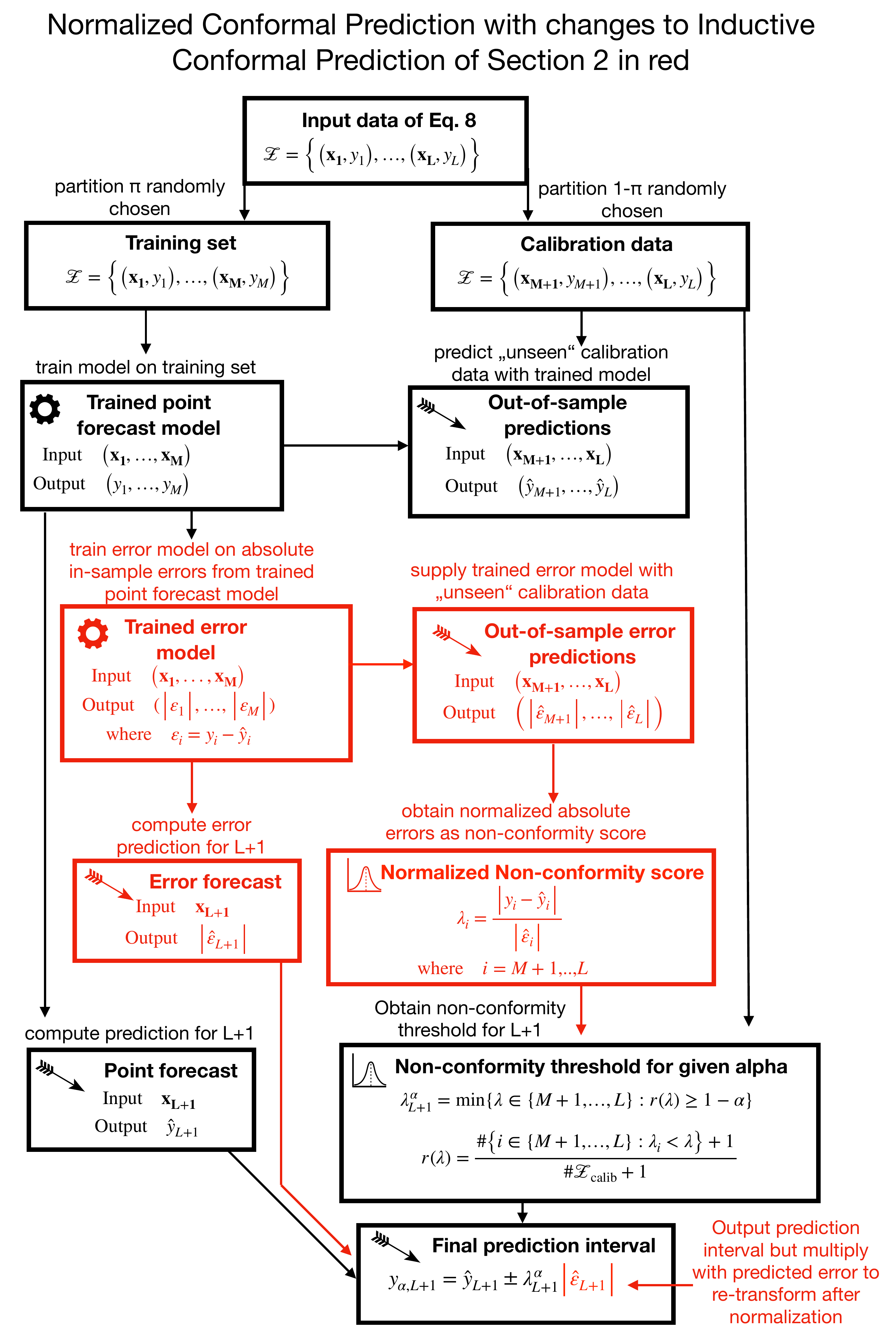}
\par\end{centering}
\caption{Schematic representation of Normalized Conformal Prediction. Detailed
information on input data per market is mention in Equation 6 as well
as section 3.1. Note that the trained error model and the trained
point forecast model are in our case kept identical in terms of tuning
parameters and algorithms, i.e., if the trained point forecast model
is a Lasso predictor, the trained error model will be a Lasso predictor
as well. However, they could also vary. The detailed hyper-parameters
per model are described in Appendix A. }

\end{figure}
The toy example is helpful in understanding the basic concept but
equally important is to fine-tune the CP approach to electricity prices.
They feature high volatility and a strong seasonality observable in
weekly and daily patterns. Weekends tend to show lower price levels
just like night hours when less electricity is needed. Therefore,
the Inductive CP model introduced in chapter 2 requires taking into
account new data and information to address the issue of heteroscedasticity.
We aim to minimize any bias by an extended Conformal Prediction scheme
referred to as Normalized Conformal Prediction (\textbf{NCP}) in \citet{papadopoulos2010neural}
that considers new data as well. Hence, we have first introduced the
basic version Inductive CP in section 2 and now present a more refined
version. Whereas the toy example only uses historical data for the
determination of non-conformity scores, the expanded version also
incorporates the information set applied for the regression model.
But what is different to the calculus mentioned in section 2? It is
mainly the non-conformity score. A non-conformity score $\lambda_{i,h}$
exists for every pair of $\left(\mathbf{x}_{i,h},y_{i,h}\right)$
in $i=M+1,...,L$. Please note that we deviate from the $t,h$ notation
and use $i$ to a) establish a connection to the examples of sub-chapter
2 and b) to highlight the different order due to sampled training
and calibration that is different from the chronological $t,h$ order.
More information on the index notations is also provided by Figure
1. The non-conformity score is given by
\begin{equation}
\lambda_{i,h}=\frac{\left|y_{i,h}-\hat{y}_{i,h}\right|}{\left|\hat{\varepsilon}_{i,h}\right|},
\end{equation}
with $\left|\hat{\varepsilon}_{i,h}\right|$ being the absolute value
of the estimated error predicted by a second, explicit error estimation
model. This section introduces NCP, Eq. (11) and (12) slightly differ
from the definition of section 2. This model predicts the estimated
error of our KNN, SVM and LASSO model for the out-of-sample data.
The interval forecast is given by
\begin{equation}
y_{\alpha,T+1,h}=\hat{y}_{T+1,h}\pm\lambda_{L+1,h}^{\alpha}\left|\hat{\varepsilon}_{L+1,h}\right|.
\end{equation}
The NCP algorithm depends on two autarkic prediction models, as shown
in Figure 7. One of them aims to deliver a point forecast referred
to as $\hat{y}_{t,h}$ in Eq. (12). It might also be regarded as a
stand-alone predictor if one disregards the Conformal Prediction framework.
In detail, the point forecast for each Conformal Prediction model
is provided by either a Lasso regression, an SVM or KNN regression.
All model-specific details and hyper-parameters are discussed in section
4.2. The point forecast models use the explanatory variables of Eq.
(6). Based on these price predictions, the errors made in the training
process are calculated. The second model uses the residuals of the
price forecasts $\hat{y}_{t,h}$ as the response and forecasts the
inaccuracy present in the actual prediction approach given as $\left|\hat{\varepsilon}_{i,h}\right|$
in Eq. (11) and Eq. (12). Once both models are trained, they equally
generate their prediction on the novel calibration dataset. So to
sum up, we have the following extensions to Inductive CP of section
2:
\begin{itemize}
\item a second point forecast model that estimates the error associated
with the prediction $\hat{y}_{t,h}$, i.e., $\left|\varepsilon_{t,h}\right|$,
\item an adjusted normalized non-conformity score mentioned in Eq. (11),
\item and finally, a new interval forecast where we multiply with the predicted
error to re-transform after normalization in Eq. (12).
\end{itemize}
Please note that we apply identical models for both the price point
forecast and the error point forecast and use 75\% of all available
data points for training and 25\% for calibration of the NCP intervals.
This means that the KNN\_NCP approach uses the exact same model set-up
for both point forecasts as shown in Figure 2, i.e., all hyper-parameters
of section 4.2 as well as the explanatory variables of Eq. (6) are
used. The interested reader might also refer to the research data
in Appendix B where we present a dedicated R-markdown file that introduces
the connected R code step by step and allows for the reproduction
of the results.

\section{Empirical results}

\label{sec:5}

\subsection{General performance metrics }

\begin{table*}[t]
\centering{}\includegraphics[scale=0.66]{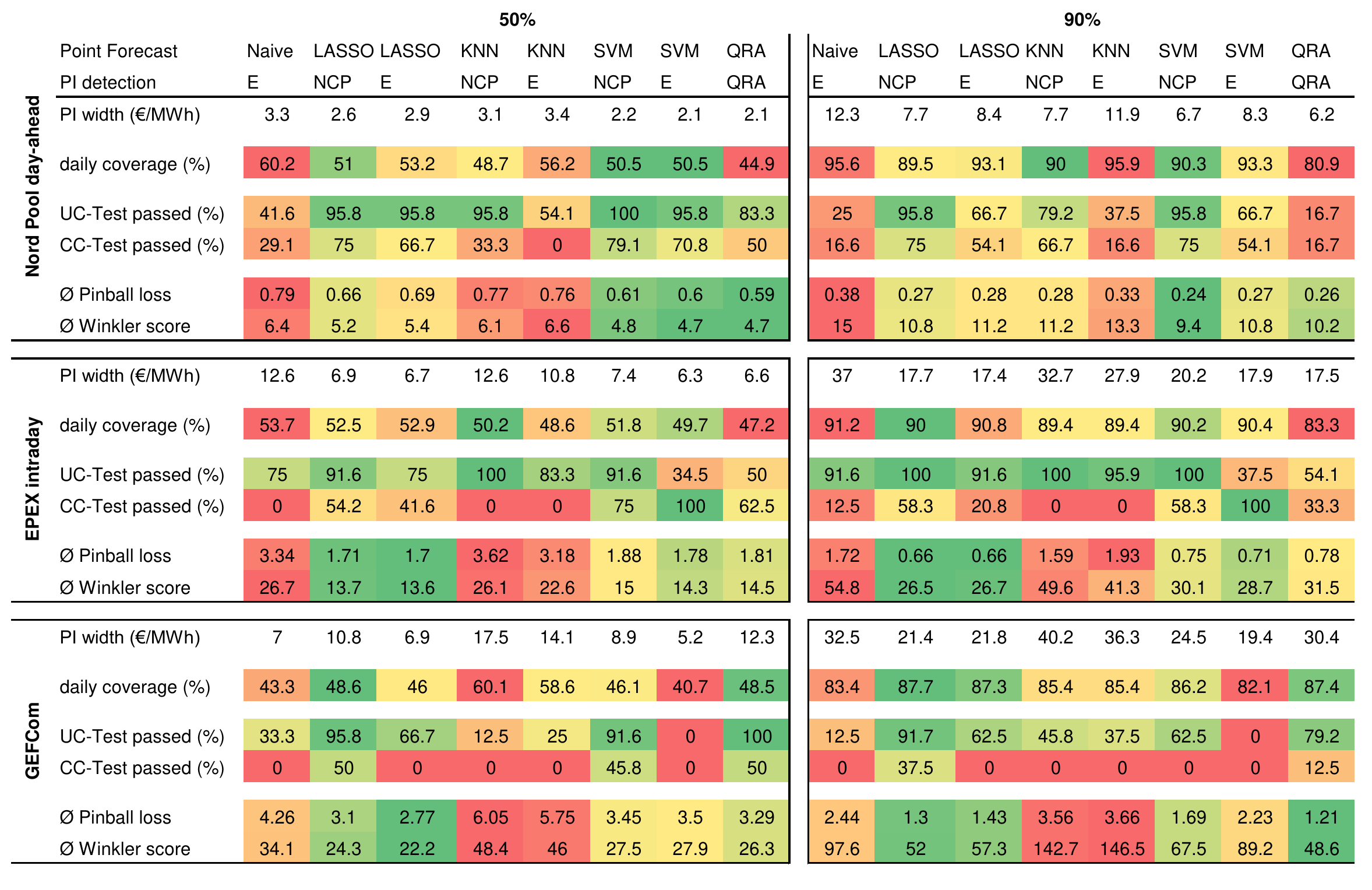}\caption{Selected prediction interval sharpness and reliability results for
empirical two-sided prediction intervals. Please note that the pinball
loss is a metrics for each quantile which we have averaged for the
respective PI, such that the 90\% PI describes the average Pinball
loss of the 5th and 95th quantile.}
\end{table*}
\begin{figure}[t]
\centering{}\includegraphics[scale=0.46]{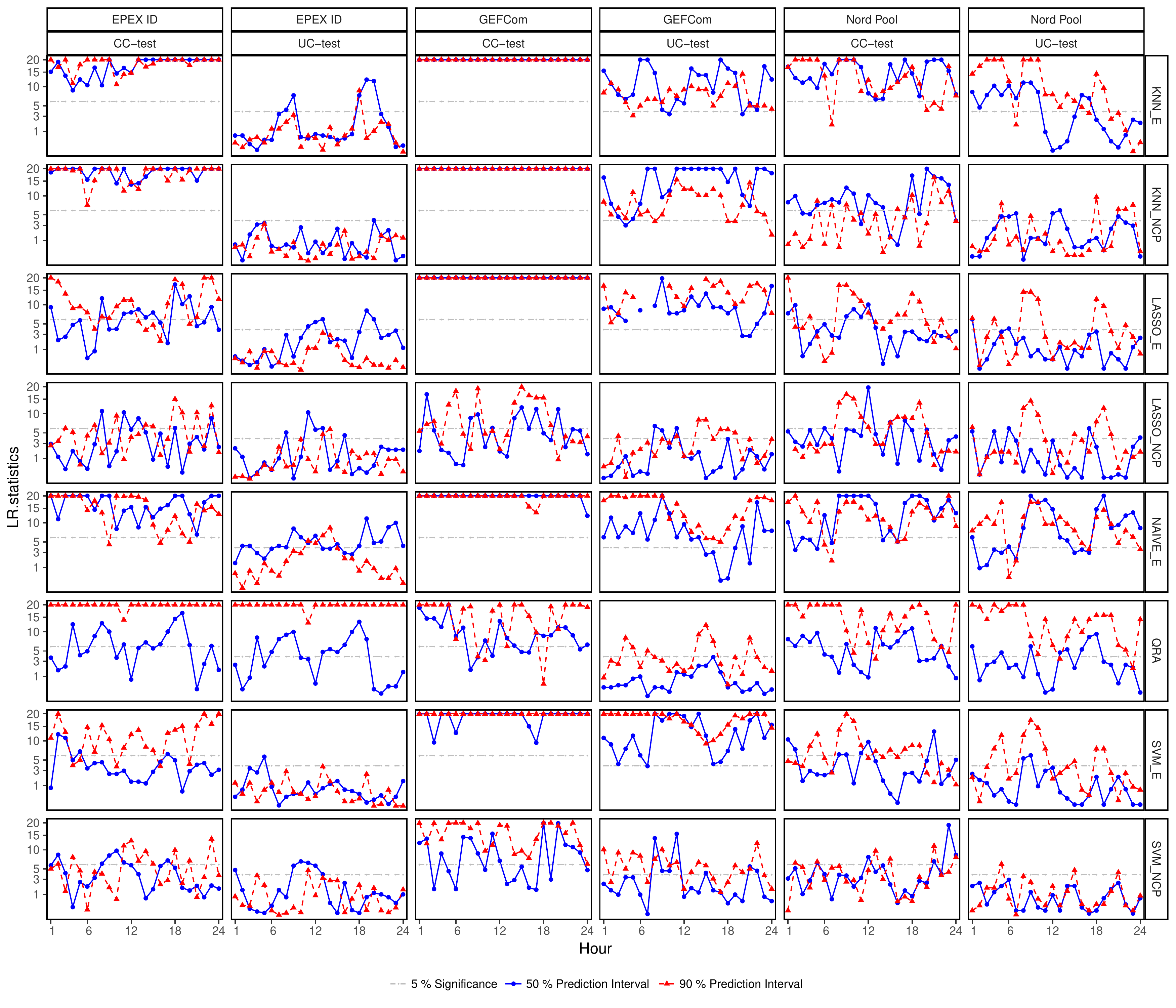}\caption{Christoffersen unconditional coverage (UC) and conditional coverage
(CC) test results reported as hourly Likelihood ratio (LR) statistics.
Please note that all LR values above 20 are set equal to 20 for the
graphical depiction.}
\end{figure}
Prediction intervals need to be reliable and sharp (\citet{nowotarski2017recent}).
The term reliability itself refers to the empirical coverage being
close or equal to the designated coverage level. It is also noteworthy
that reliability and sharpness share a close interdependency. The
sharper an interval gets, the less it is near the true coverage. Moreover,
we are facing a trade-off between the two criteria. As a first approach
to the topic, we compare the empirical coverage with the nominal values
under consideration of the PI width in the upper rows of Table 2.
From left to right, it depicts values for each model and interval.
The first impression is that the NCP models yield good coverage. Yet,
this is not a uniform statement as the results differ per model and
interval. There is no single best model for coverage, nor for sharpness.
If we take a closer look at the different markets, it appears as if
the EPEX intraday and GEFCom markets are more difficult to predict
in a probabilistic manner as their error measures are higher than
Nord Pool ones. This intuitively makes sense as these markets are
the more volatile ones. Higher volatility seems to widen the difference
between predictions and observations. Our QRA model shows good performance
but remains behind the Conformal Prediction models. Please note that
we can validate our QRA results by means of the findings reported
in \citet{nowotarski2014merging} for the Nord Pool market as they
are very similar. QRA results obtained in \citet{nowotarski2017recent}
for the GEFCom dataset were slightly better than our QRA model which
might be due to the changed selection of point forecast models. We
chose our predictors mostly out of the field of machine learning,
while the aforementioned authors used a wider set of traditional time
series approaches. However, since the results do not fundamentally
differ, we see that as further cross-literature validation of our
models.\\
\hspace*{0.5cm}A downside of the previous analysis is the strict
focus on both the 50\% and 90\% prediction intervals and their associated
25/75 and 5/95 percentiles. This enforces symmetry and does not evaluate
the upper and lower parts of the PI in a separate way, which leaves
room for netting effects in errors. In order to assess the prediction
quality, one needs to focus on all other percentiles, as done in Figure
6. It depicts the deviation between empirical and nominal coverage
computed for all percentiles in steps of 5 and shows the asymmetric
estimation quality. The first striking fact is that, contrary to Table
2, the Nord Pool and GEFCom markets appear to be harder to predict
since the distance to the true coverage is higher than anticipated
by Table 2. Most of the models seem to suffer around the 55 and 45
percentile which is usually a difficult region to predict due to the
high density of observations in that area. We did not compute the
50 percentile as this is typically estimated by median point forecasts
and is not directly associated with the Conformal Prediction technique
anymore. For reasons of a clear depiction, the 50 percentile area
was only interpolated. There is no single best predictor for all different
markets. Support vector machines tend to show a constant level of
differences in comparison with other estimation approaches. If we
compare the empirical quantiles with their NCP equivalents it is not
possible to favor one over the other. The choice of the best model
seems to be heavily connected with the market to be predicted and
the underlying point forecast. Our last finding is associated with
the observation of differences between Table 2 and Figure 6. Obviously,
singular percentiles are harder to foretell. But if one, for instance,
considers the QRA performance in the EPEX intraday market, an interesting
relationship becomes evident. The deviation switches from positive
to very negative. If we recall that the 50\% PI should cover the range
of the 25 and 75 quantiles, we might assume that some of the models
benefit from netting effects out of symmetry. This explains why the
Nord Pool study reveals higher deviations in Figure 6. The 50\% PI
is near to the nominal coverage but the two individual quantiles are
less close.

\subsection{Christoffersen test}

Besides the nominal coverage, there is a commonly used test set provided
by \citet{christoffersen1998evaluating} which examines unconditional
coverage (UC), independence, and conditional coverage (CC). We stick
to \citet{weron2008forecasting} and restrict on the first observation
which renders conditional coverage to be the sum of independence and
unconditional coverage. That being said, it is sufficient to test
for unconditional coverage and its conditional equivalent as the latter
comprises the independence information. The tests are processed in
the Likelihood-Ratio (LR) framework and use a hit series (1 if the
interval is correct, 0 otherwise) as input. We also test hourly observations
so that no daily effects can falsely create any signals of dependence
across several hours. We use the R-package \texttt{rugarch} with
its 'VaRTest' function to compute results. The detailed test statistics
are displayed in Figure 8. We plot the test output in the form of
LR test statistics against each hour of the day and do so per model
and market. The dashed gray line reflects the 5\% significance level
of the test statistics and determines the acceptance criterion for
both the UC and CC test. All statistics above the gray line point
towards a lack of reliability under our test setup. The first striking
observation in Figure 8 is that Conformal Prediction appears to have
a positive effect on the LR statistics. Taking the Lasso, for instance,
most of the NCP plots per market are below their empirical counterparts
which speaks for the theoretical foundation that postulates true coverage
for NCP. We can also observe such performance for SVM\_NCP predictions.
KNN leaves a mixed impression. It seems to have consistent problems
in all markets with the stricter CC test in particular. QRA's 50\%
PI values are mostly under the 5\% significance level. The associated
90\% PI test statistics create a different impression as they mostly
do not meet our acceptance criterion. These findings are in line with
\citet{nowotarski2017recent} in case of GEFCom but partially differ
in the Nord Pool case. \citet{nowotarski2014merging} report a higher
ratio of accepted hours for the 90\% PI. Still, this might be caused
by the different blend of point forecasts. Finally, Naive\_E requires
a deeper look. Our naive benchmark yields insufficient reliability
in all of the considered power markets which delivers evidence to
the fact that a more advanced prediction interval determination approach
brings additional benefit. If we compare the intra-market results,
it becomes evident that GEFCom is the time series with the worst reliability,
even for the most performant models. In contrast to that, EPEX ID
and Nord Pool seem to have roughly the same range of errors across
all models. Unfortunately, we do not know enough about the GEFCom
origin to establish any further connection between fundamental characteristics
and the problems with reliability. Yet, we can acknowledge that out
of our three time series the Christoffersen test confirms that GEFCom
is the most difficult to estimate.

\subsection{Winkler Score and pinball loss}

\begin{figure}[t]
\begin{centering}
\includegraphics[scale=0.55]{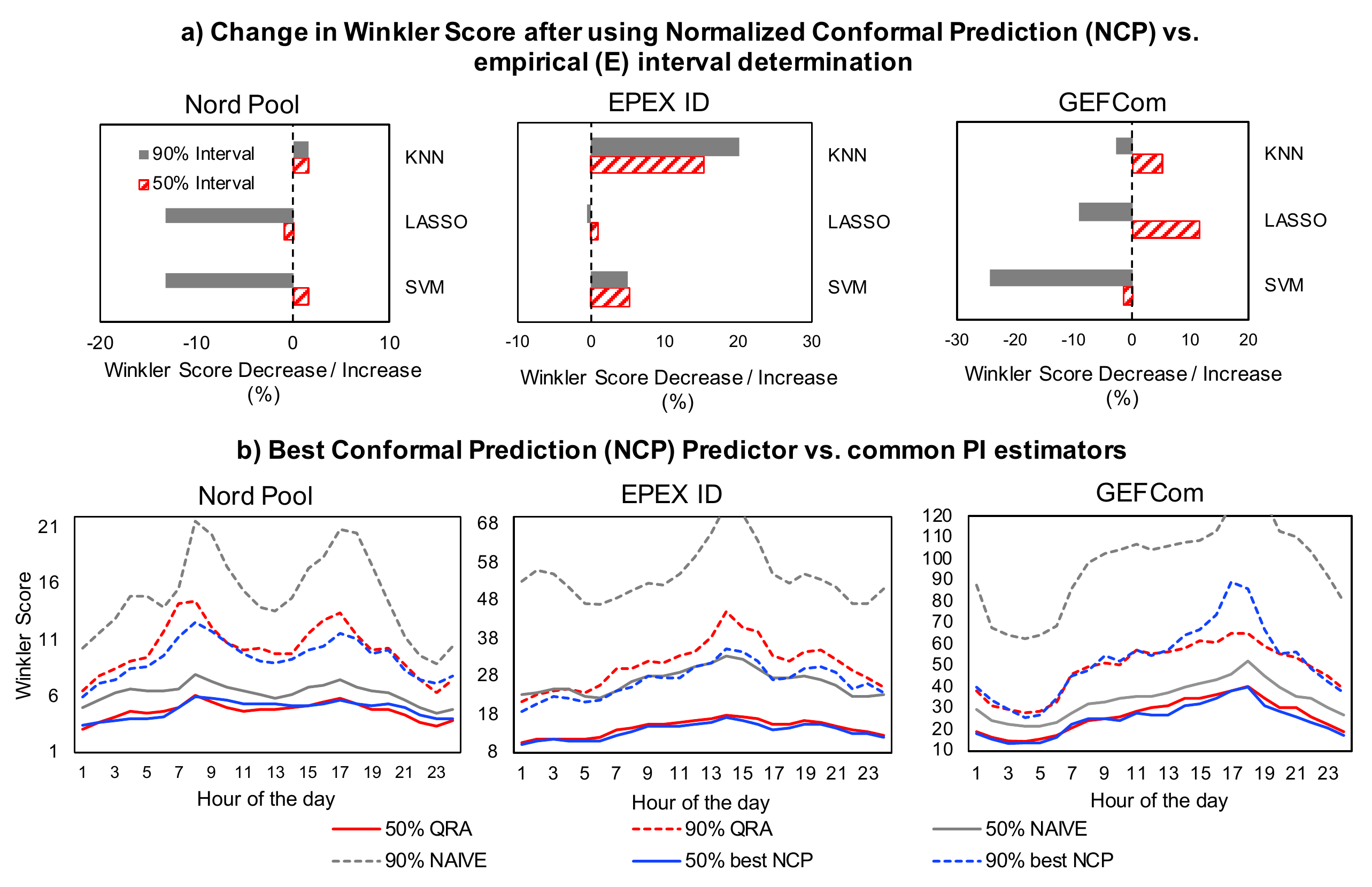}
\par\end{centering}
\caption{Model performance measured by the Winkler Score as introduced in Eq.
(13). Part a) compares the NCP PI determination with the empirical
error distribution (E) approach per forecast model (e.g. SVM\_NCP
vs. SVM\_E) to identify the benefits of Conformal Prediction PIs,
while section b) sets the best Conformal Prediction method based on
the Winkler Score reported in Table 2 in relation to the naive benchmark
and QRA. In detail, the best models are LASSO\_NCP for EPEX ID and
GEFCom, and SVM\_NCP in the case of Nord Pool data.}
\end{figure}
All previous assessments have focused either on reliability or sharpness
in a separate manner. A metric known as the Winkler Score (see \citet{winkler1972decision}
for the derivation) allows for the joint elicitation of both, given
in (cases representation adopted from \citet{maciejowska2016probabilistic})
\begin{equation}
W_{t,h}=\begin{cases}
\textrm{B}_{t,h} & \textrm{for}\,y_{t,h}\text{\ensuremath{\in}}[\textrm{L}_{t,h},\textrm{U}_{t,h}]\\
\textrm{B}_{t,h}+\frac{2}{\alpha}(\textrm{L}_{t,h}-y_{t,h}) & \textrm{for}\,y_{t,h}<\mathbf{\textrm{L}_{\mathit{t,h}}}\\
\textrm{B}_{t,h}+\frac{2}{\alpha}(y_{t,h}-\textrm{U}_{t,h}) & \textrm{for}\,y_{t,h}>\mathbf{\textrm{U}_{\mathit{t,h}}},
\end{cases}
\end{equation}
where $\textrm{B}_{t,h}$ represents the width of the two-sided prediction
interval, i.e., $\textrm{B}_{t,h}=\textrm{U}_{t,h}-\textrm{L}_{t,h}$,
and $\textrm{L}_{t,h},\textrm{U}_{t,h}$ its lower and upper bounds.
The Winkler Score penalizes deviating coverage and examines the width.
All results are depicted in Figure 9. The upper part under section
a) tries to contribute to the question of additional benefits of using
Normalized Conformal Prediction in combination with different point
forecasts. Which point forecast models gain the most from Normalized
Conformal Prediction and consequently feature the lowest Winkler Score?
Figure 9 shows the decrease in the latter if we use NCP instead of
the error distribution approach (using an \_NCP model instead of an
\_E one). For GEFCom and Nord Pool, one can observe a decrease in
the Winkler Score of about 10\% - 20\%, occurring mostly with the
90\% PIs marked in gray. Interestingly, the 50\% PIs increase in most
of the cases. Hence, the forecast performance in the mid quantiles
suffers from NCP. In EPEX intraday markets, NCP additions do not have
a positive impact on the Winkler Score which underlines our diverse
choice of markets and how different the results are. The KNN model
even shows an increase in the German intraday market albeit for all
other markets there is at least a bit of decrease in the error measure.
A possible connection could be established to Figure 6 where the EPEX
ID market features low deviation from the true coverage. If we recall
that the Winkler Score takes into account coverage we might assume
that this market is overall less complex in its prediction characteristics
and, therefore, does not benefit from further model complexity. Still,
this is just a first, trivial explanation and requires more empirical
analysis that goes beyond the scope of this paper. All in all, Normalized
Conformal Prediction seems to have a positive impact on the Winkler
Score in two of the three markets. On the other hand, the performance
varies with the underlying point forecast model even in the same market.\\
\hspace*{0.5cm} Section a) gives a good first impression but leaves
the time structure of a short-term price forecast aside. We want to
assess hourly differences and have plotted a corresponding curve of
hourly Winkler Scores for each market in section b). In order to reduce
the complexity of the graphical depiction, we have narrowed down the
analysis and only compare the best Normalized Conformal Prediction
model (LASSO\_NCP for EPEX ID and GEFCom, SVM\_NCP in case of Nord
Pool data) based on Table 2 with QRA and our naive benchmark. Not
surprisingly, the Winkler Score curve of the naive approach is much
higher, which implies less accuracy. This holds true for all three
markets. QRA and NCP are very close: in our Nord Pool and EPEX intraday
application, NCP features slightly lower curves while with GEFCom
data, QRA and NCP are almost equal. If one takes a deeper look at
the hourly shape of each individual curve it becomes evident that
night hours show lower Winkler Scores. There are spikes in the error
measure during the off-peak/peak time block shifts (around hour 8
and 20) in the Nord Pool market. This effect is often observed in
electricity spot markets or day-ahead markets in particular and might
be explained by additional power plants ramped up or down to cover
peak load during the day. While intraday markets are usually used
to cover residual loads or renewables adjustments, day-ahead markets
serve as a market place for much larger volumes. Therefore, we observe
a strong block shifting effect in the Nord Pool day-ahead data while
there is less in the intraday equivalent. Taking the hourly shapes
into account, we have to favor QRA or NCP over the naive benchmark,
with NCP showing a slightly lower Winkler Score in some instances.
\\
\hspace*{0.5cm}Our second test statistic is a very popular one. The
pinball loss (PB loss) was chosen to be the official scoring rule
for the GEFCom 2014 probabilistic forecasting track in \citet{hong2016probabilistic}
and gained the reputation of a common measure for probabilistic forecasts.
Its representation is given by

\begin{equation}
PB(q_{y_{t,h}}(\tau),y_{t,h})=\begin{cases}
(1-\tau)(q_{y_{t,h}}(\tau)-y_{t,h}) & \textrm{for}\,y_{t,h}<q_{y_{t,h}}(\tau)\\
\tau(y_{t,h}-q_{y_{t,h}}(\tau)) & \textrm{for}\,y_{t,h}\geq q_{y_{t,h}}(\tau),
\end{cases}
\end{equation}
where $(q_{y_{t,h}}(\tau)$ is the $\tau$-th estimated quantile of
the electricity price series $y_{t,h}$. The pinball loss is a quantile
specific measure but can simply be averaged across hours or quantiles
in order to have a more comprehensive sharpness indicator. The analysis
of the pinball loss goes in a different direction compared with the
Winkler assessments since it focuses on percentiles in order to determine
how an approach behaves under varying probabilistic assumptions. This
modus operandi also shifts the focus towards asymmetric performance
and sets each percentile in a performance relation. In contrast to
that, Table 2 focuses on prediction intervals which imply symmetry.
All findings are presented in Figure 10. The first thing that has
to be noted is the difference in scale. In comparison with the Nord
Pool market, the EPEX intraday and GEFCom plots comprise 5 or 9 times
higher PB loss scores. This corresponds to the previous impression
we had from Table 2 or the numeric values of the Winkler analysis
where these markets were more difficult to predict as well. All in
all, the conclusion drawn from Figure 10 is similar to the one in
Figure 6. The middle percentiles increase the error measure. But there
is another connection to this expression. All models except the KNN
and the naive one are very close in terms of performance. Yet, there
is one pattern. Normalized Conformal Prediction suffers in the middle
percentiles to an extent that the much simpler \_E models have a lower
PB loss. The picture changes once the outer percentiles are concerned.
If we recall the results of Table 2, NCP models were yielding, in
general, a bit better coverage and PI width. On the other hand, both
the 90\% and 50\% PI only consider the 5/95 or 25/75 quantile, respectively.
The picture seems to differ with more median oriented quantiles. Depending
on the market, the best performing model can either be NCP, QRA or
an error distribution approach which reflects that there is no single
best predictor when it comes to the PB loss.

\begin{figure}[!t]
\begin{centering}
\includegraphics[scale=0.55]{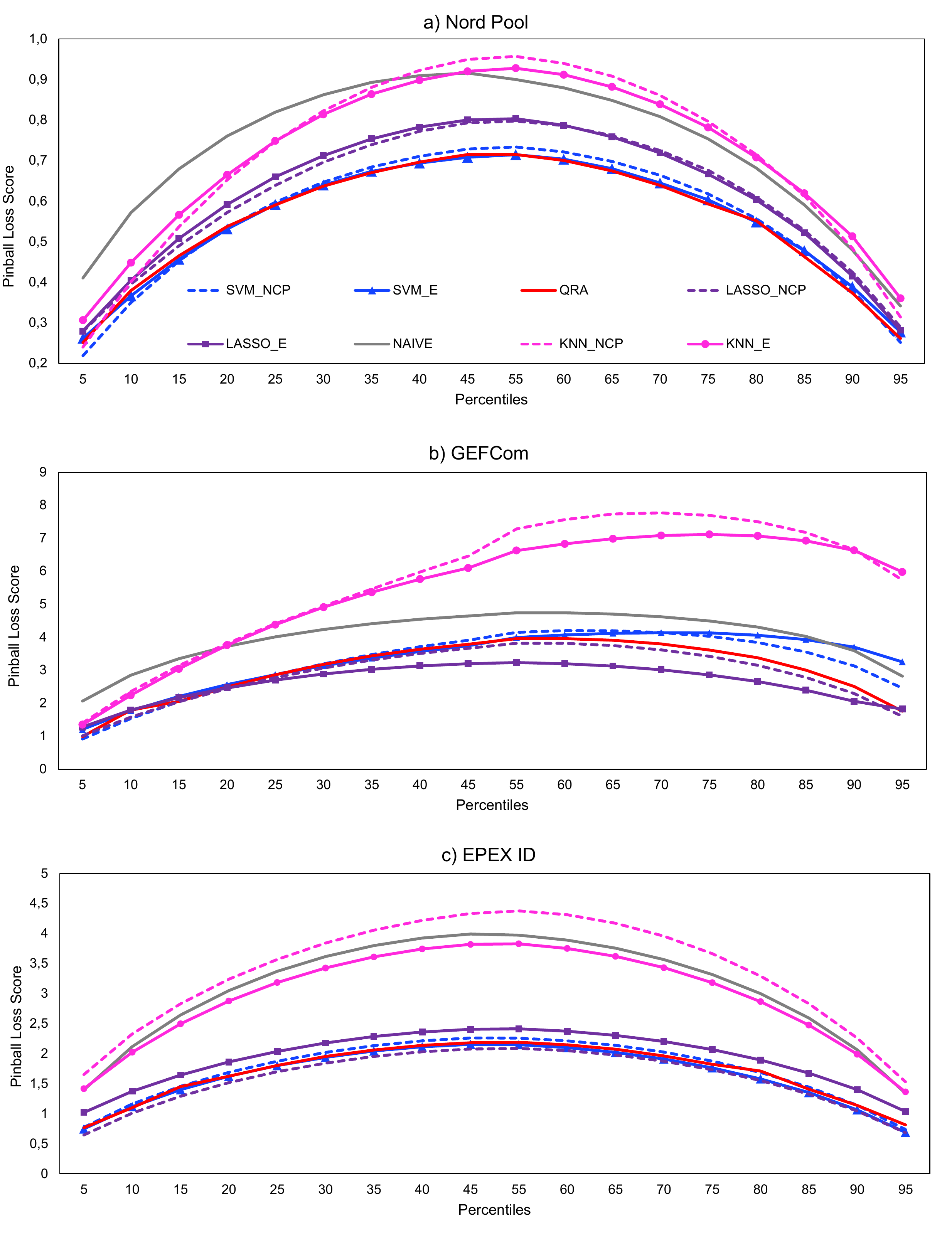}
\par\end{centering}
\caption{Pinball loss scores per percentile as mentioned in Eq. (14). Please
note that we have taken the average over all 24 hours to depict model
performance under different percentiles and have plotted every 5th
percentile except the 50th one.}
\end{figure}

\subsection{Path dependent evaluation of Conformal Prediction performance drivers}

The previous sub-chapters have only taken a global view on estimation
capabilities and compared the model performance with QRA and a naive
benchmark. We have not discussed the question of why Conformal Prediction
is performing in a decent manner. Leaving the technical concepts aside,
Conformal Prediction features three possible origins from which performance
might stem. Firstly, it forces the forecast to be symmetric. We sort
the non-conformity measure $\lambda_{i,h}$ and consider the respective
value corresponding to the desired PI. The identified non-conformity
score is added or subtracted from the forecast such that there is
no designated differentiation between quantiles. For instance, we
determine the 50\% PI by subtracting and adding the same $\lambda_{i,h}$
from our point forecast. In contrast to that, an asymmetric approach
determines the 25th and 75th quantile in an independent manner. Combining
these two quantiles yields the 50\% PI in a second step. The second
potential source of performance gains is the sampling technique described
in chapter two. Conformal Prediction randomly splits the available
set of information into training and calibration to ensure a maximum
of generalization. But does this step really improve the models? The
third aspect of Conformal Prediction is, at least intuitively, a very
important one. In the case of Normalized Conformal Prediction, we
adjust the non-conformity score by estimated errors as mentioned in
Eq. (12). When it comes to the forecast value $\hat{y}_{t+1,h}$,
this small modification ensures that all new information $\mathbf{x}_{t+1,h}$
is regarded in the prediction interval determination by firstly estimating
the error for t+1 and then plugging it in in Eq. (12). Without any
quantitative backing, one will surely assume that this is a reasonable
operation with a positive impact on predictive performance, especially
if we consider the strong daily effects of electricity price time
series. Heteroscedasticity caused by weekly effects is taken into
account since we include the daily dummy in the new information set.\\
\hspace*{0.5cm}We run a simulation path of different combinations
of the above three model expansions that jointly form the core of
Conformal Prediction in the same out-of-sample fashion that was already
applied in the empirical analysis in the previous sub-chapters. The
regression model of Eq. (6), all hyper-parameters discussed in section
4.2, the transformations, and the out-of-sample rolling window approach
remain unchanged. At the same time, we only consider the 3 point forecast
models KNN, SVM and Lasso which were used for the previous Conformal
Prediction computations. The following models were evaluated in the
analysis path:
\begin{itemize}
\item \textbf{asymmetric quantiles}: We compute quantiles of non-absolute
errors $\varepsilon_{t,h}=y_{t,h}-\hat{y}_{t,h}$ where $\hat{y}_{t,h}$
stems from KNN, Lasso and SVM point forecasts trained on all available
data points (i.e., the training and calibration set). Since the error
distribution is almost symmetric, we can simply compute the lower
bound as a quantile of the negative errors (where the point forecast
underestimated the price) and the upper bound as the quantile of the
positive errors. The bounds are then added to or subtracted from the
point forecast $\hat{y}_{t,h}.$
\item \textbf{quantiles - sampled}: The model is identical to the asymmetric
quantiles model besides the fact that we randomly sample 75\% of the
available data points (i.e., the training and calibration set) for
training of the point forecasts and use the remaining 25\% calibration
set for the calculation of quantiles of errors $\varepsilon_{i,h}=y_{i,h}-\hat{y}_{i,h}$.
Hence, this approach avoids overfitting, as discussed in section 2.
\item \textbf{quantiles - normalized}: The model is almost identical to
the asymmetric quantiles model but the basis for the quantile calculation
$\varepsilon_{t,h}=y_{t,h}-\hat{y}_{t,h}$ is normalized with the
expected error $\hat{\varepsilon}_{t,h}$ produced by a second point
forecast model (again either SVM, KNN or LASSO) using the explanatory
variables of Eq. (6) to forecast the errors. We use absolute values
of the estimated error $\hat{\varepsilon}_{t,h}$ to prevent negative
values to change the entire prediction to be negative. Hence, we compute
the quantiles of $\frac{(y_{t,h}-\hat{y}_{t,h})}{\left|\hat{\varepsilon}_{t,h}\right|}$
and multiply the final PI forecasts with $\left|\hat{\varepsilon}_{t,h}\right|$
to upscale the values again. Please note that no sampling or split
into calibration and training is applied. Instead, we use in-sample
errors to train the second model which yields the expected error $\hat{\varepsilon}_{t,h}$.
\item \textbf{quantiles - symmetric}: This model is identical to the empirical
error distribution approach of section 4.3.1 (denoted as \_E models)
and uses absolute errors for the computation of quantiles.
\item \textbf{Conformal Prediction}: The CP model is the same as the one
discussed in the toy example in section 2. The only difference to
the NCP of section 4.3.3 is the more simple, non-normalized computation
of the non-conformity score such that $\lambda_{i,h}=\left|y_{i,h}-\hat{y}_{i,h}\right|$.
\item \textbf{quantiles - norm-sampled}: Assume the model described under
'quantiles-normalized', i.e., normalized quantiles calculated from
errors $\varepsilon_{i,h}=(y_{i,h}-\hat{y}_{i,h})$ and then normalized
by means of $\left|\hat{\varepsilon}_{i,h}\right|$ with an additional
75\% / 25\% random sampling of the data. The training period is not
fully exploited here to avoid overfitting.
\item \textbf{quantiles norm-symmetric}: The norm-symmetric model is the
same as the Conformal Prediction one but does not use any sampling.
It can also be seen as the toy example of section 2 except for the
sample split. We just use in-sample errors to derive the non-conformity
scores.
\item \textbf{Normalized Conformal Prediction}: This is the model of section
4.3.2.
\end{itemize}
Speaking of values, we utilize our two common error measures, PB loss
and Winkler Score, to identify the impact of normalization, sampling
and symmetry. In addition, we compare the PI width. Although this
is not a traditional error measure per se, it helps in understanding
differences. Figures 11 and 12 illustrate both the different simulation
paths as well as the connected error measures separated into the 50\%
and 90\% prediction interval. We start the analysis with the most
basic form of PI estimation by computing the quantiles of the empirical
error distribution\footnote{Please note that this differs from the \_E model used before due to
the lack of symmetry. All \_E models are symmetric ones.}, depicted at the very left. This model neither samples any of the
data nor uses new information. Please also note that we receive an
asymmetric estimation as we independently compute the quantile for
the upper and lower part of the PI, i.e., do not use absolute errors
for quantile determination. We initially assumed that this PI predictor
is by far the worst one but were proven wrong by our empirical study.
In comparison to Normalized Conformal Prediction, the asymmetric empirical
quantiles tend to perform very well. In the German intraday market,
the results are nearly equal to NCP, while NCP yields lower errors
in Nord Pool and GEFCom. But as a first result, we can say that the
most modest form of probabilistic forecasting is more accurate than
expected.\\
\hspace*{0.5cm}Based on the asymmetric quantiles, we separately add
all three extension stages to the basic model. The second estimator
uses normalization via estimated errors, and its other two equivalents
add symmetry and sampling of data. Performance-wise, one can observe
a clear picture. Adding symmetry lowers the Winkler Score and PB loss.
Interestingly, the interval width is not widened at the same time
which reflects that we yield more accurate intervals and that the
previous one was not just too narrow. Such a clear indication came
unanticipated as the technical model difference is rather small. Instead
of stand-alone quantiles we compute absolute empirical errors and
add or subtract them from the point forecast. While this is a very
small change in terms of computation, its impact is impressive and
speaks for symmetry in residuals. \\
\hspace*{0.5cm}Sampling to avoid overfitting appears to further increase
accuracy, which at least partially refutes our asymptotic argument
of making no mistake without sampling. Still, the effect is very small.
Normalization or the addition of new information is a bit problematic
in a stand-alone application. The asymmetric empirical model computes
negative and positive normalization values in the form of estimated
errors to be subtracted or added to the probabilistic forecast. These
values are then normalized by the expected error, which is either
a positive or negative value. It might occur that the sign of some
of the values changes the entire PI to unrealistic estimations which
is why we adjust the normalization numbers to be always positive.
All in all, the computation of normalized asymmetric quantiles does
not really make sense. However, in the interest of completeness, we
show the model results. In some cases, we observe a tremendous performance
drop after normalization, which further underlines the argument of
a known misconstruction.\\
\hspace*{0.5cm}We could end our analysis at this point. But that
would imply linear additivity of the specific model extensions. Is
it intuitively possible to add, for instance, symmetry and normalization
and yield the sum of each extension's performance? We expand our models
into three different paths to answer this question. Firstly, we add
the three other extensions. The normalized empirical quantiles are
changed to symmetric ones. This step shall further validate our findings
with regards to the inefficiency of non-symmetric normalization. And
indeed, symmetry solves the issue of misconstructed PIs. Error measures
are lower while the PI width is narrowed down as well, which reflects
an improvement of sharpness under reliability.\\
\hspace*{0.5cm}A different picture is painted if we extend the sampled
asymmetric quantiles to a symmetric estimator. Please note that this
predictor is the same as the Conformal Prediction mentioned in chapter
two. The Winkler Score and PB loss only slightly change in some instances,
which highlights that sampling helps on a case by case basis. In contrast
to that, adding normalization to sampled quantiles causes the same
bias as with the normalized quantiles. Due to the lack of symmetry,
the sign of the output could change the entire prediction, which causes
the Winkler Score and PB loss to be much higher. Hence, our empirical
study suggests that the path from empirical quantiles to normalized
and then normalized and sampled ones does not make sense to apply.
\\
\hspace*{0.5cm}Last but not least, we focus on Normalized Conformal
Prediction as our last layer. In some cases, such as the GEFCom predictions,
it makes sense to utilize all three extensions jointly. In other scenarios,
such as EPEX intraday, the addition of sampling to norm-symmetric
intervals was not beneficial with regards to performance. This finding
perfectly matches the impression from the Winkler Score analysis.
The less complex markets with regards to estimations, namely Nord
Pool and EPEX ID, do not seem to benefit from model extension in the
way the GEFCom data set does. \\
\hspace*{0.5cm}
\begin{figure}
\begin{centering}
\includegraphics[scale=0.52]{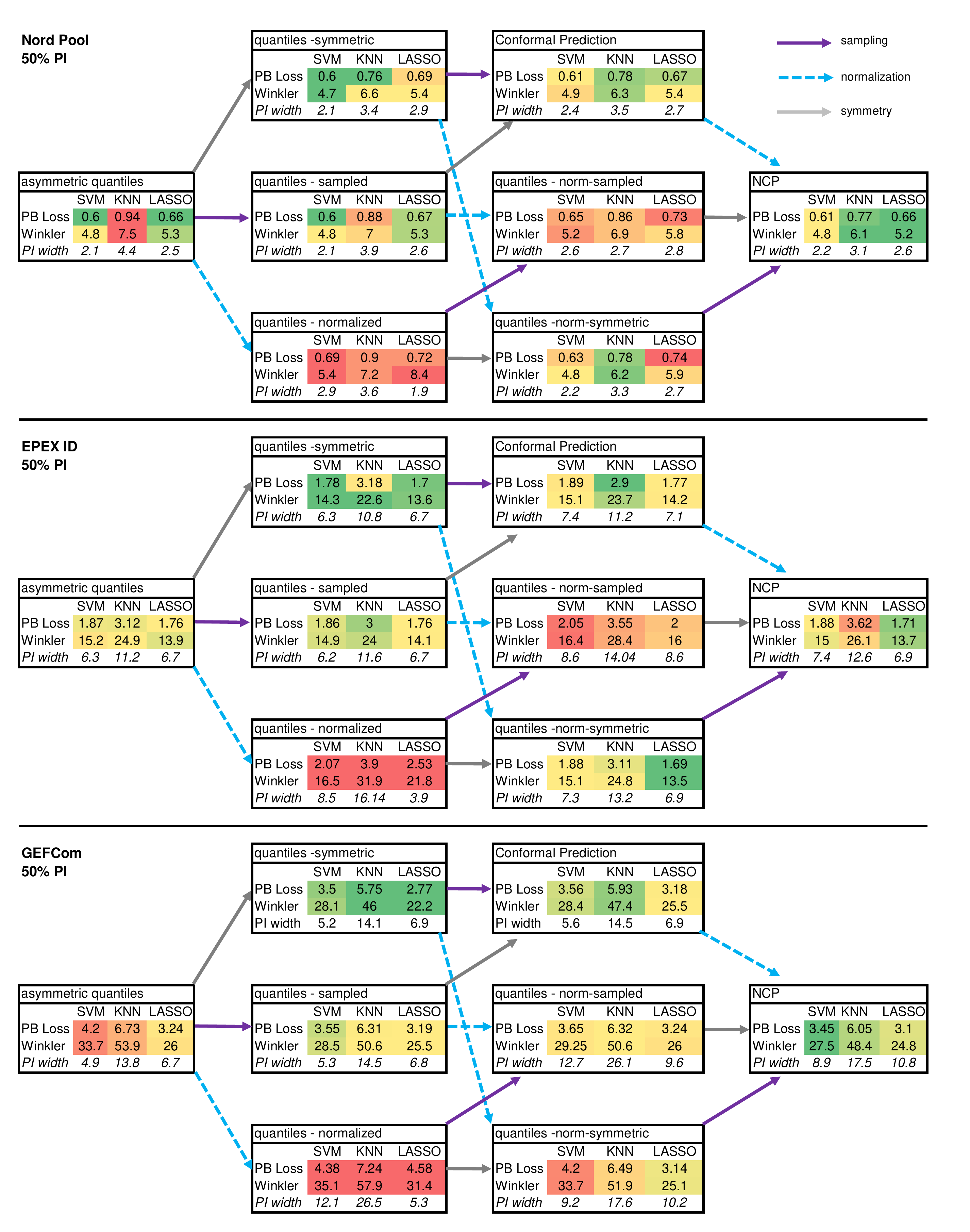}\caption{Identification of Conformal Prediction's key performance drivers based
on a path dependent analysis of the three different model extensions,
symmetry, normalization and sampling for the 50\% PI. We applied the
same empirical setup as described in the previous sub-chapters but
changed the models bit by bit (each addition is depicted by the colored
arrows) to evaluate which part of Conformal Prediction accounts for
most of the gains in precision. Please note that the pinball loss
is a metric for each quantile which we averaged for the respective
PI, such that the 50\% PI describes the average pinball loss of the
25th and 75th quantile. The heat-map colors only compare PI models,
i.e., we compare KNN asymmetric quantiles with KNN under normalized
quantiles but do not compare KNN versus SVM performance.}
\par\end{centering}
\end{figure}
\begin{figure}
\begin{centering}
\includegraphics[scale=0.52]{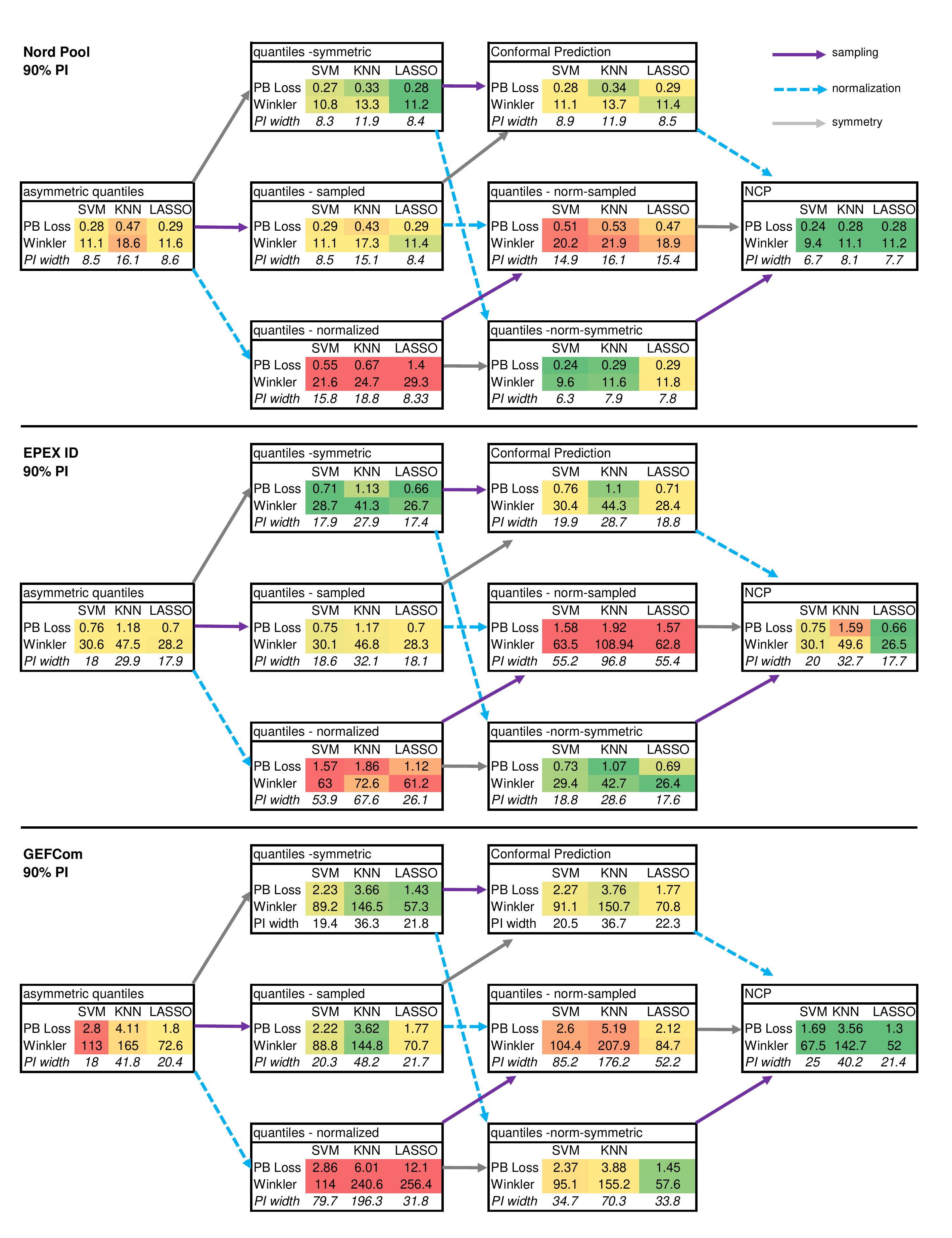}\caption{Identification of Conformal Prediction's key performance drivers based
on a path dependent analysis of the three different model extensions,
symmetry, normalization and sampling for the 90 \% PI. The plot is
equivalent to Figure 11 besides the different PI. We extend the models
bit by bit (each addition is depicted by the colored arrows) and show
the impact on the displayed metrics. The heat-map colors only compare
PI models, i.e., we compare KNN asymmetric quantiles with KNN under
normalized quantiles but do not compare KNN versus SVM performance
as the analysis shall focus on the performance improvement that is
due to different PI computations.}
\par\end{centering}
\end{figure}
So, after computing 144 models, what does the path-dependent analysis
suggest? We can assume that there is no singular model that outperforms
all the others. Our choice of markets is a very diverse one which
causes results to be different. The same counts for the point prediction
models themselves. There are some universal tendencies, such as beneficial
effects of symmetric estimations. That being said, the usage of normalization
and sampling only adds value in some of the cases. We advise every
forecaster to carefully test the probabilistic models in question,
especially if the market to be predicted features statistical similarity
to the GEFCom data. Conformal Prediction serves as a good framework
but still requires fine-tuning with regards to the optimal blend of
its key components.

\section{Verdict and possible value-chain implications for energy traders}

\label{sec:6}

\subsection{Conclusion and outlook}

The underlying research motivation of this paper was a thorough introduction
of Conformal Prediction with a particular focus on short-term electricity
prices. We have discussed the theoretical concept and demonstrated
that Conformal Prediction works like a second layer to any given point
forecast. By exploiting errors made from these point forecasts, symmetric
prediction intervals are computed. The other two novelties in that
sense are the sampling which ensures a high level of generalization
and the normalization by means of estimated errors. We explicitly
consider new information when we adjust the PI with an estimated error
for $t+1$. This helps to account for electricity price characteristics
like heteroscedasticity caused by contrasting load scenarios on different
days since we include information about such in the probabilistic
estimation process. Leaving the theory behind, we test multiple probabilistic
forecasting concepts in three independent pricing regimes and establish
a connection to the empirical results of \citet{nowotarski2014merging}
and \citet{nowotarski2017recent} by adopting a comparable QRA model.
We demonstrate that Conformal Prediction can live up to the expectations
and yields valid prediction intervals even with changing point forecast
inputs. In comparison with a naive benchmark, the well-known QRA and
a simple error distribution approach applied to a similar set of point
forecasts, NCP is equal or even better in terms of Winkler Score,
Christoffersen test or PB loss. Connected to the decent performance
is the question of key performance drivers. An additional evaluation
that independently analyzes Conformal Prediction's three key aspects,
symmetry, normalization and sampling of input data, brings more clarity.
We simulated different paths leading to a total of 144 computations.
The overall picture is rather unclear. Conformal Prediction and its
different modifications show varying performance across markets. As
a consequence, we advise energy companies to compute a path like the
one in sub-chapter 5.4 for their forecasting problems and only then
decide on one specific Conformal Prediction model.\\
\hspace*{0.5cm}We have to acknowledge that these findings only apply
to short-term electricity prices. We deliberately chose to focus on
these in order to yield maximum objectivity in our analysis. Future
research might look at other possible applications, such as wind forecasting
or load prediction. Apart from that, we did not discuss any extension
of the known Normalized Conformal Prediction framework. Following
the idea of QRA, Conformal Prediction intervals might also be averaged
to get even better results. However, current research has only started
to focus on aggregated Conformal Prediction but the first findings
are promising. The interested reader might take a closer look at \citet{carlsson2014aggregated}
for a description of multiple aggregated Conformal Predictors or \citet{vovk2015cross}
for the idea of combining various p-values or non-conformity scores.
It might be worth expanding these ideas to the world of energy-related
forecasting.

\subsection{Benefits for power traders}

Besides gaining academic knowledge, is there any practical implication
that companies active in the field of energy trading could consider
for their value-chain? We believe that the model-agnostic character
together with the post-processing aspect could be a benefit in energy
trading. Assume a power producer with a thermal generation unit for
instance. Depending on the technical specification, most units feature
a minimum runtime such that instantaneous changes in their steering
are hard to realize (see \citet{petersen2014heuristic} for an example
on how runtimes are considered in the underlying optimization problem).
Suppose a minimum runtime of three hours. The unit output is sold
on the day-ahead market with limit orders depending on the marginal
costs of the plant. Bidding in day-ahead markets is easier for traders
according to \citet{weber2010adequate}. It allows them to consider
technical restrictions like the three-hour minimum runtime using linked
orders also known as block-orders, which are not present in continuous
intraday trading. In the continuous intraday market, it is possible
to buy power back that has already been sold day-ahead and turn off
the unit, which then saves fuel costs and leaves the spread between
the day-ahead market and intraday trading as profit. \citet{kiesel2017econometric}
also mention the start-up costs of thermal units that additionally
impede real-time steering based on price changes or different load
scenarios in the intraday market. Referring to our three hour minimum
runtime example, we could buy back if the current price level for
the three consecutive hours is lower than the production costs of
the unit. However, this is not ideal as it may be the case that the
first hour is already below production costs and the other two hours
will be later, due to the continuous nature of intraday trading. All
three hours may be below the production costs at a different point
in time. If we only consider situations where all three hours are
jointly below production costs, commercial opportunities will remain
unused.\\
\hspace*{0.5cm}A forecast is needed in this scenario to assess the
likelihood of each hour being below the production costs in later
trading stages. Most companies will use a point forecast to do so.
The problem is that forecasts imply uncertainty. If the company trades
based on a simple point estimation, it does not consider the accuracy
of the forecast. A Conformal Prediction-based model could be added
to the existing point forecast without any greater computational effort
and would allow for easy determination of PIs. We have shown possibilities
to compute them in an out-of-sample manner to ensure generalization
and tackle heteroscedasticity. If this is done, the power producer
can trade based on PIs. The PIs and their confidence level are automatically
valid and shall be selected depending on the risk appetite of each
company. While the point forecast only takes into account the mean
or median, it might make sense to trade on the 50\% PI and evaluate
later hours based on larger intervals since their longer trading time
naturally implies more risk. All in all, a probabilistic forecast
allows for more accurate trading decisions and a greater likelihood
of executed intraday trades, i.e., higher power plant revenues as
also proposed by \citet{amjady2006energy}. The great benefit of Conformal
Prediction is that it is accurate without requiring much effort, which
is why we think that it is perfectly suited to companies that have
one existing, proven point forecast and want to fine-tune their trading
decisions based on uncertainty and individual risk acceptance.

\section*{Acknowledgements}

We thank the participants of the S3 seminar on 08.11.2018 in Wroclaw,
Poland for their comments and questions that greatly improved the
manuscript. This research did not receive any specific grant from
funding agencies in the public, commercial or not-for-profit sectors.
This research article was partially supported by the German Research
Foundation (DFG,Germany) and the National Science Center (NCN, Poland)
through BEETHOVEN grantno. 2016/23/G/HS4/01005.

\appendix

\section*{Appendix A: Detailed Description of Point Forecast Models}

\subsection*{Naive expert learner}

A simple model is required to assess if more sophisticated approaches
truly add any benefit. Therefore, we assume that the best guess for
today's price is the last available similar day price. Based on the
scheme laid out in \citet{nowotarski2017recent} we use yesterday's
hourly price if the day to be predicted is a Tuesday, Wednesday, Thursday
or Friday. If not, then the price of the hour of the previous week
is assumed to be the forecast. This \textbf{Naive }benchmark\textbf{
}does not require any computations nor transformations but regards
weekly effects and the daily term-structure due to its multivariate
approach.

\subsection*{Lasso regression}

Our second expert learner combines point forecasting with feature
selection. Introduced in \citet{tibshirani1996regression}, the least
absolute shrinkage and selection operator (\textbf{Lasso}) enhances
the common ordinary least squares (OLS) scheme so that unnecessary
variables are penalized or even removed. The Lasso estimator expands
OLS by adding a linear penalty factor $\zeta_{t,h}\geq0$ in
\begin{equation}
\hat{\mathbf{\mathbf{\mathrm{\boldsymbol{\beta}}}}}_{\mathbf{lasso}}=\argmin_{\mathbf{\boldsymbol{\beta}}_{\mathit{\mathbf{h}}}}\left(\underbrace{\sum_{t=1}^{T}(y_{t,h}-\sum_{j=1}^{p}\beta_{j,h}x_{t,j,h})^{2}}_{\mathrm{\textrm{RSS}}}+\underbrace{\zeta_{t,h}\sum_{j=1}^{p}\left|\beta_{j,h}\right|}_{\mathrm{\textrm{Penalty Term}}}\right),
\end{equation}
where $x_{t,j,h}$ are the Yeo-Johnson transformed explanatory variables
mentioned in Eq. (6). Note that the lasso penalty is often denoted
by $\lambda$ but since this parameter is already used for the non-conformity
score, the lasso penalty is called $\zeta_{t,h}$ in this paper. In
case of $\zeta_{t,h}=0$ we obtain OLS results, while $\zeta_{t,h}\rightarrow\infty$
causes all variables to be removed from the model. We compute a solution
for $\hat{\mathbf{\mathbf{\mathrm{\boldsymbol{\beta}}}}}_{\mathbf{lasso}}$
using the coordinate descent algorithm implemented in the R package
\texttt{glmnet} of \citet{Friedman}. The algorithm itself leaves
the hyper-parameter $\zeta_{t,h}$ to be optimized. We use a two-fold
cross-validation and identify the ideal tuning parameter each hour
and day out of an equidistant grid between 0.1 and 0.001 with a step
size 0.001. Although this results in more computational effort, a
recent study in \citet{uniejewski2018efficient} highlights the importance
of recursive Lasso hyper-parameter tuning and its beneficial effect
on performance.

\subsection*{K-nearest neighbor regression}

The idea of the K-nearest neighbor (denoted as\textbf{ KNN}) algorithm
is based on the fact that patterns in data will repeat in the future.
The model implies that a comparable set of input factors will most
likely result in the same output as observed with analogous input
factors. Therefore, KNN approaches use a similarity measure to identify
observations with similar patterns. The most similar values are then
regarded as the prediction. The parameter $k$ defines how many similar
observations are taken into account. If $k>1$, the different realizations
for the target variable are usually averaged in order to yield an
estimate for the true value. We explicitly incorporate this rather
simple approach to have an alternative estimator based on a simple
mapping rule. We use $k=50$ for Nord Pool and GEFCom data and $k=200$
for EPEX intraday prices and determine similarity based on Euclidean
distance. The KNN model is trained with all the explanatory variables
of Eq. (6). The computations are done applying the R-package \texttt{FNN}
with the additional setting 'algorithm=c(\textquotedbl kd\_tree\textquotedbl ,
\textquotedbl cover\_tree\textquotedbl , \textquotedbl brute\textquotedbl )'.
Hence, we do not preliminary chose an algorithm but select the one
that has the best in-sample accuracy and apply that for out-of-sample
predictions. We use untransformed prices for the KNN calculation as
transformations do not add any benefit. For another application of
KNN on Spanish day-ahead prices, the interested reader might refer
to \citet{lora2002comparison}.

\subsection*{Support vector machine regression}

A support vector machine regression (\textbf{SVM})\footnote{Please note that we use SVM as an abbreviation for the regression
case of support vector machines, also denoted as support vector regression.} maps the regression data to a high-dimensional space and tries to
find simple linear decision rules in a new space. Its foundation is
given by simple geometric principles such as maximum margin hyperplane.
From a computational point of view, we can consider SVM as a convex
optimization. Its original application was limited to classification
problems but later generalized to deal with regression applications
(see \citet{vapnik1997support} for instance). The solution obtained
is a global one. A large variety of kernel functions renders SVM models
be very flexible, which is why most of their practical applications
are in the context of hybrid models that combine several model layers
together. The choice of kernels also controls the ability to capture
non-linear problems or solely linear ones. Typical examples of applications
are to be found in \citet{che2010short} or \citet{zhang2012day}.
In contrast to that, we apply a simple stand-alone model based on
the R package \texttt{kernlab} that uses a radial basis function
kernel which is nonlinear. The explanatory variables supplied to the
SVM learner are the ones mentioned in Eq. (6). We set $sigma=0.005$
and apply a cost of constraint violation of $C=1.25$. The algorithm
itself is restricted to a maximum of 1,000 iterations and works on
Yeo-Johnson transformed prices.

\section*{Appendix B: Research data}

Supplementary data to this article, such as a dedicated R-markdown
file of the Conformal Prediction algorithm, can be found online for
the sake of full reproducibility at \href{https://data.mendeley.com/datasets/3wnk2pz6y2/1}{DOI:10.17632/3wnk2pz6y2.1}.

\renewcommand*{\bibfont}{\footnotesize}
\begingroup \raggedright \sloppy

\bibliographystyle{model5-names}
\nocite{*}
\bibliography{Literature}

\begin{thebibliography}{51}
\expandafter\ifx\csname natexlab\endcsname\relax\def\natexlab#1{#1}\fi
\providecommand{\url}[1]{\texttt{#1}}
\providecommand{\href}[2]{#2}
\providecommand{\path}[1]{#1}
\providecommand{\DOIprefix}{doi:}
\providecommand{\ArXivprefix}{arXiv:}
\providecommand{\URLprefix}{URL: }
\providecommand{\Pubmedprefix}{pmid:}
\providecommand{\doi}[1]{\href{http://dx.doi.org/#1}{\path{#1}}}
\providecommand{\Pubmed}[1]{\href{pmid:#1}{\path{#1}}}
\providecommand{\bibinfo}[2]{#2}
\ifx\xfnm\relax \def\xfnm[#1]{\unskip,\space#1}\fi
\bibitem[{Aggarwal et~al.(2009)Aggarwal, Saini \&
  Kumar}]{aggarwal2009electricity}
\bibinfo{author}{Aggarwal, S.~K.}, \bibinfo{author}{Saini, L.~M.}, \&
  \bibinfo{author}{Kumar, A.} (\bibinfo{year}{2009}).
\newblock \bibinfo{title}{Electricity price forecasting in deregulated markets:
  A review and evaluation}.
\newblock {\it \bibinfo{journal}{International Journal of Electrical Power \&
  Energy Systems}\/},  {\it \bibinfo{volume}{31}\/}, \bibinfo{pages}{13--22}.
\bibitem[{Aitkin(1996)}]{aitkin1996general}
\bibinfo{author}{Aitkin, M.} (\bibinfo{year}{1996}).
\newblock \bibinfo{title}{A general maximum likelihood analysis of
  overdispersion in generalized linear models}.
\newblock {\it \bibinfo{journal}{Statistics and computing}\/},  {\it
  \bibinfo{volume}{6}\/}, \bibinfo{pages}{251--262}.
\bibitem[{Amjady \& Hemmati(2006)}]{amjady2006energy}
\bibinfo{author}{Amjady, N.}, \& \bibinfo{author}{Hemmati, M.}
  (\bibinfo{year}{2006}).
\newblock \bibinfo{title}{Energy price forecasting-problems and proposals for
  such predictions}.
\newblock {\it \bibinfo{journal}{IEEE Power and Energy Magazine}\/},  {\it
  \bibinfo{volume}{4}\/}, \bibinfo{pages}{20--29}.
\bibitem[{Bordignon et~al.(2013)Bordignon, Bunn, Lisi \&
  Nan}]{bordignon2013combining}
\bibinfo{author}{Bordignon, S.}, \bibinfo{author}{Bunn, D.~W.},
  \bibinfo{author}{Lisi, F.}, \& \bibinfo{author}{Nan, F.}
  (\bibinfo{year}{2013}).
\newblock \bibinfo{title}{Combining day-ahead forecasts for british electricity
  prices}.
\newblock {\it \bibinfo{journal}{Energy Economics}\/},  {\it
  \bibinfo{volume}{35}\/}, \bibinfo{pages}{88--103}.
\bibitem[{Breiman(2001)}]{breiman2001random}
\bibinfo{author}{Breiman, L.} (\bibinfo{year}{2001}).
\newblock \bibinfo{title}{Random forests}.
\newblock {\it \bibinfo{journal}{Machine learning}\/},  {\it
  \bibinfo{volume}{45}\/}, \bibinfo{pages}{5--32}.
\bibitem[{Bunn et~al.(2013)Bunn, Andresen, Chen, Westgaard
  et~al.}]{bunn2013analysis}
\bibinfo{author}{Bunn, D.}, \bibinfo{author}{Andresen, A.},
  \bibinfo{author}{Chen, D.}, \bibinfo{author}{Westgaard, S.} et~al.
  (\bibinfo{year}{2013}).
\newblock \bibinfo{title}{Analysis and forecasting of electricity price risks
  with quantile factor models}.
\newblock In {\it \bibinfo{booktitle}{Finance Research Seminar Series,
  University of St. Gallen}\/}.
\bibitem[{Buuren \& Groothuis-Oudshoorn(2011)}]{buuren2011mice}
\bibinfo{author}{Buuren, S.}, \& \bibinfo{author}{Groothuis-Oudshoorn, K.}
  (\bibinfo{year}{2011}).
\newblock \bibinfo{title}{mice: Multivariate imputation by chained equations in
  {R}}.
\newblock {\it \bibinfo{journal}{Journal of statistical software}\/},  {\it
  \bibinfo{volume}{45}\/}, \bibinfo{pages}{1--67}.
\bibitem[{Carlsson et~al.(2014)Carlsson, Eklund \&
  Norinder}]{carlsson2014aggregated}
\bibinfo{author}{Carlsson, L.}, \bibinfo{author}{Eklund, M.}, \&
  \bibinfo{author}{Norinder, U.} (\bibinfo{year}{2014}).
\newblock \bibinfo{title}{Aggregated conformal prediction}.
\newblock In {\it \bibinfo{booktitle}{IFIP International Conference on
  Artificial Intelligence Applications and Innovations}\/} (pp.
  \bibinfo{pages}{231--240}).
\newblock \bibinfo{organization}{Springer}.
\bibitem[{Che \& Wang(2010)}]{che2010short}
\bibinfo{author}{Che, J.}, \& \bibinfo{author}{Wang, J.}
  (\bibinfo{year}{2010}).
\newblock \bibinfo{title}{Short-term electricity prices forecasting based on
  support vector regression and auto-regressive integrated moving average
  modeling}.
\newblock {\it \bibinfo{journal}{Energy Conversion and Management}\/},  {\it
  \bibinfo{volume}{51}\/}, \bibinfo{pages}{1911--1917}.
\bibitem[{Christoffersen(1998)}]{christoffersen1998evaluating}
\bibinfo{author}{Christoffersen, P.~F.} (\bibinfo{year}{1998}).
\newblock \bibinfo{title}{Evaluating interval forecasts}.
\newblock {\it \bibinfo{journal}{International economic review}\/},  {\it
  \bibinfo{volume}{39}\/}, \bibinfo{pages}{841--862}.
\bibitem[{Dudek(2016)}]{dudek2016multilayer}
\bibinfo{author}{Dudek, G.} (\bibinfo{year}{2016}).
\newblock \bibinfo{title}{Multilayer perceptron for gefcom2014 probabilistic
  electricity price forecasting}.
\newblock {\it \bibinfo{journal}{International Journal of Forecasting}\/},
  {\it \bibinfo{volume}{32}\/}, \bibinfo{pages}{1057--1060}.
\bibitem[{Friedman et~al.(2010)Friedman, Hastie \& Tibshirani}]{Friedman}
\bibinfo{author}{Friedman, J.}, \bibinfo{author}{Hastie, T.}, \&
  \bibinfo{author}{Tibshirani, R.} (\bibinfo{year}{2010}).
\newblock \bibinfo{title}{Regularization paths for generalized linear models
  via coordinate descent.}
\newblock {\it \bibinfo{journal}{Journal of Statistical Software}\/},  {\it
  \bibinfo{volume}{33}\/}, \bibinfo{pages}{1 -- 22}.
\bibitem[{Gammerman et~al.(1998)Gammerman, Vovk \&
  Vapnik}]{gammerman1998learning}
\bibinfo{author}{Gammerman, A.}, \bibinfo{author}{Vovk, V.}, \&
  \bibinfo{author}{Vapnik, V.} (\bibinfo{year}{1998}).
\newblock \bibinfo{title}{Learning by transduction}.
\newblock In {\it \bibinfo{booktitle}{Proceedings of the Fourteenth conference
  on Uncertainty in artificial intelligence}\/} (pp.
  \bibinfo{pages}{148--155}).
\newblock \bibinfo{organization}{Morgan Kaufmann Publishers Inc.}
\bibitem[{Hoaglin(2003)}]{hoaglin2003john}
\bibinfo{author}{Hoaglin, D.~C.} (\bibinfo{year}{2003}).
\newblock \bibinfo{title}{John w. tukey and data analysis}.
\newblock {\it \bibinfo{journal}{Statistical Science}\/},  {\it
  \bibinfo{volume}{18}\/}, \bibinfo{pages}{311--318}.
\bibitem[{Hong et~al.(2016)Hong, Pinson, Fan, Zareipour, Troccoli \&
  Hyndman}]{hong2016probabilistic}
\bibinfo{author}{Hong, T.}, \bibinfo{author}{Pinson, P.}, \bibinfo{author}{Fan,
  S.}, \bibinfo{author}{Zareipour, H.}, \bibinfo{author}{Troccoli, A.}, \&
  \bibinfo{author}{Hyndman, R.~J.} (\bibinfo{year}{2016}).
\newblock \bibinfo{title}{Probabilistic energy forecasting: Global energy
  forecasting competition 2014 and beyond}.
\newblock {\it \bibinfo{journal}{International Journal of Forecasting}\/},
  {\it \bibinfo{volume}{32}\/}, \bibinfo{pages}{896--913}.
\bibitem[{Hubicka et~al.(2018)Hubicka, Marcjasz \& Weron}]{hubicka2018note}
\bibinfo{author}{Hubicka, K.}, \bibinfo{author}{Marcjasz, G.}, \&
  \bibinfo{author}{Weron, R.} (\bibinfo{year}{2018}).
\newblock \bibinfo{title}{A note on averaging day-ahead electricity price
  forecasts across calibration windows}.
\newblock {\it \bibinfo{journal}{IEEE Transactions on Sustainable Energy}\/},
  {\it \bibinfo{volume}{10}\/}, \bibinfo{pages}{321--323}.
\bibitem[{Johansson et~al.(2014)Johansson, Bostr{\"o}m, L{\"o}fstr{\"o}m \&
  Linusson}]{johansson2014regression}
\bibinfo{author}{Johansson, U.}, \bibinfo{author}{Bostr{\"o}m, H.},
  \bibinfo{author}{L{\"o}fstr{\"o}m, T.}, \& \bibinfo{author}{Linusson, H.}
  (\bibinfo{year}{2014}).
\newblock \bibinfo{title}{Regression conformal prediction with random forests}.
\newblock {\it \bibinfo{journal}{Machine learning}\/},  {\it
  \bibinfo{volume}{97}\/}, \bibinfo{pages}{155--176}.
\bibitem[{Keles et~al.(2016)Keles, Scelle, Paraschiv \&
  Fichtner}]{keles2016extended}
\bibinfo{author}{Keles, D.}, \bibinfo{author}{Scelle, J.},
  \bibinfo{author}{Paraschiv, F.}, \& \bibinfo{author}{Fichtner, W.}
  (\bibinfo{year}{2016}).
\newblock \bibinfo{title}{Extended forecast methods for day-ahead electricity
  spot prices applying artificial neural networks}.
\newblock {\it \bibinfo{journal}{Applied energy}\/},  {\it
  \bibinfo{volume}{162}\/}, \bibinfo{pages}{218--230}.
\bibitem[{Khosravi et~al.(2013)Khosravi, Nahavandi \&
  Creighton}]{khosravi2013neural}
\bibinfo{author}{Khosravi, A.}, \bibinfo{author}{Nahavandi, S.}, \&
  \bibinfo{author}{Creighton, D.} (\bibinfo{year}{2013}).
\newblock \bibinfo{title}{A neural network-garch-based method for construction
  of prediction intervals}.
\newblock {\it \bibinfo{journal}{Electric Power Systems Research}\/},  {\it
  \bibinfo{volume}{96}\/}, \bibinfo{pages}{185--193}.
\bibitem[{Kiesel \& Paraschiv(2017)}]{kiesel2017econometric}
\bibinfo{author}{Kiesel, R.}, \& \bibinfo{author}{Paraschiv, F.}
  (\bibinfo{year}{2017}).
\newblock \bibinfo{title}{Econometric analysis of 15-minute intraday
  electricity prices}.
\newblock {\it \bibinfo{journal}{Energy Economics}\/},  {\it
  \bibinfo{volume}{64}\/}, \bibinfo{pages}{77--90}.
\bibitem[{Kowalczewski(2019)}]{kowalczewski2019normalized}
\bibinfo{author}{Kowalczewski, J.} (\bibinfo{year}{2019}).
\newblock \bibinfo{title}{Normalized conformalprediction for time series data}.
\newblock \bibinfo{howpublished}{accessed online 1st July, 2020}.
\newblock \URLprefix \url{http://oa.upm.es/57817/1/TFM_JAKUB_KOWALCZEWSKI.pdf}.
\bibitem[{Liaw \& Wiener(2002)}]{rf}
\bibinfo{author}{Liaw, A.}, \& \bibinfo{author}{Wiener, M.}
  (\bibinfo{year}{2002}).
\newblock \bibinfo{title}{Classification and regression by randomforest}.
\newblock {\it \bibinfo{journal}{R News}\/},  {\it \bibinfo{volume}{2}\/},
  \bibinfo{pages}{18--22}. \URLprefix
  \url{http://CRAN.R-project.org/doc/Rnews/}.
\bibitem[{Liu et~al.(2015)Liu, Nowotarski, Hong \&
  Weron}]{liu2015probabilistic}
\bibinfo{author}{Liu, B.}, \bibinfo{author}{Nowotarski, J.},
  \bibinfo{author}{Hong, T.}, \& \bibinfo{author}{Weron, R.}
  (\bibinfo{year}{2015}).
\newblock \bibinfo{title}{Probabilistic load forecasting via quantile
  regression averaging on sister forecasts}.
\newblock {\it \bibinfo{journal}{IEEE Transactions on Smart Grid}\/},  {\it
  \bibinfo{volume}{8}\/}, \bibinfo{pages}{730--737}.
\bibitem[{Lora et~al.(2002)Lora, Santos, Santos, Exp{\'o}sito \&
  Ramos}]{lora2002comparison}
\bibinfo{author}{Lora, A.~T.}, \bibinfo{author}{Santos, J.~R.},
  \bibinfo{author}{Santos, J.~R.}, \bibinfo{author}{Exp{\'o}sito, A.~G.}, \&
  \bibinfo{author}{Ramos, J. L.~M.} (\bibinfo{year}{2002}).
\newblock \bibinfo{title}{A comparison of two techniques for next-day
  electricity price forecasting}.
\newblock In {\it \bibinfo{booktitle}{International Conference on Intelligent
  Data Engineering and Automated Learning}\/} (pp. \bibinfo{pages}{384--390}).
\newblock \bibinfo{organization}{Springer}.
\bibitem[{Maciejowska \& Nowotarski(2016)}]{maciejowska2016hybrid}
\bibinfo{author}{Maciejowska, K.}, \& \bibinfo{author}{Nowotarski, J.}
  (\bibinfo{year}{2016}).
\newblock \bibinfo{title}{A hybrid model for gefcom2014 probabilistic
  electricity price forecasting}.
\newblock {\it \bibinfo{journal}{International Journal of Forecasting}\/},
  {\it \bibinfo{volume}{32}\/}, \bibinfo{pages}{1051--1056}.
\bibitem[{Maciejowska et~al.(2016)Maciejowska, Nowotarski \&
  Weron}]{maciejowska2016probabilistic}
\bibinfo{author}{Maciejowska, K.}, \bibinfo{author}{Nowotarski, J.}, \&
  \bibinfo{author}{Weron, R.} (\bibinfo{year}{2016}).
\newblock \bibinfo{title}{Probabilistic forecasting of electricity spot prices
  using factor quantile regression averaging}.
\newblock {\it \bibinfo{journal}{International Journal of Forecasting}\/},
  {\it \bibinfo{volume}{32}\/}, \bibinfo{pages}{957--965}.
\bibitem[{Marcjasz et~al.(2018)Marcjasz, Serafin \&
  Weron}]{marcjasz2018selection}
\bibinfo{author}{Marcjasz, G.}, \bibinfo{author}{Serafin, T.}, \&
  \bibinfo{author}{Weron, R.} (\bibinfo{year}{2018}).
\newblock \bibinfo{title}{Selection of calibration windows for day-ahead
  electricity price forecasting}.
\newblock {\it \bibinfo{journal}{Energies}\/},  {\it \bibinfo{volume}{11}\/},
  \bibinfo{pages}{2364}.
\bibitem[{Nowotarski et~al.(2014)Nowotarski, Raviv, Tr{\"u}ck \&
  Weron}]{nowotarski2014empirical}
\bibinfo{author}{Nowotarski, J.}, \bibinfo{author}{Raviv, E.},
  \bibinfo{author}{Tr{\"u}ck, S.}, \& \bibinfo{author}{Weron, R.}
  (\bibinfo{year}{2014}).
\newblock \bibinfo{title}{An empirical comparison of alternative schemes for
  combining electricity spot price forecasts}.
\newblock {\it \bibinfo{journal}{Energy Economics}\/},  {\it
  \bibinfo{volume}{46}\/}, \bibinfo{pages}{395--412}.
\bibitem[{Nowotarski \& Weron(2014)}]{nowotarski2014merging}
\bibinfo{author}{Nowotarski, J.}, \& \bibinfo{author}{Weron, R.}
  (\bibinfo{year}{2014}).
\newblock \bibinfo{title}{Merging quantile regression with forecast averaging
  to obtain more accurate interval forecasts of nord pool spot prices}.
\newblock In {\it \bibinfo{booktitle}{European Energy Market (EEM), 2014 11th
  International Conference on the}\/} (pp. \bibinfo{pages}{1--5}).
\newblock \bibinfo{organization}{IEEE}.
\bibitem[{Nowotarski \& Weron(2015)}]{nowotarski2015computing}
\bibinfo{author}{Nowotarski, J.}, \& \bibinfo{author}{Weron, R.}
  (\bibinfo{year}{2015}).
\newblock \bibinfo{title}{Computing electricity spot price prediction intervals
  using quantile regression and forecast averaging}.
\newblock {\it \bibinfo{journal}{Computational Statistics}\/},  {\it
  \bibinfo{volume}{30}\/}, \bibinfo{pages}{791--803}.
\bibitem[{Nowotarski \& Weron(2018)}]{nowotarski2017recent}
\bibinfo{author}{Nowotarski, J.}, \& \bibinfo{author}{Weron, R.}
  (\bibinfo{year}{2018}).
\newblock \bibinfo{title}{Recent advances in electricity price forecasting: A
  review of probabilistic forecasting}.
\newblock {\it \bibinfo{journal}{Renewable and Sustainable Energy Reviews}\/},
  {\it \bibinfo{volume}{81}\/}, \bibinfo{pages}{1548--1568}.
\bibitem[{Papadopoulos \& Haralambous(2010)}]{papadopoulos2010neural}
\bibinfo{author}{Papadopoulos, H.}, \& \bibinfo{author}{Haralambous, H.}
  (\bibinfo{year}{2010}).
\newblock \bibinfo{title}{Neural networks regression inductive conformal
  predictor and its application to total electron content prediction}.
\newblock In {\it \bibinfo{booktitle}{International Conference on Artificial
  Neural Networks}\/} (pp. \bibinfo{pages}{32--41}).
\newblock \bibinfo{organization}{Springer}.
\bibitem[{Petersen et~al.(2014)Petersen, Hansen, Bendtsen, Edlund \&
  Stoustrup}]{petersen2014heuristic}
\bibinfo{author}{Petersen, M.~K.}, \bibinfo{author}{Hansen, L.~H.},
  \bibinfo{author}{Bendtsen, J.}, \bibinfo{author}{Edlund, K.}, \&
  \bibinfo{author}{Stoustrup, J.} (\bibinfo{year}{2014}).
\newblock \bibinfo{title}{Heuristic optimization for the discrete virtual power
  plant dispatch problem}.
\newblock {\it \bibinfo{journal}{IEEE Transactions on Smart Grid}\/},  {\it
  \bibinfo{volume}{5}\/}, \bibinfo{pages}{2910--2918}.
\bibitem[{Raviv et~al.(2015)Raviv, Bouwman \& van Dijk}]{raviv2015forecasting}
\bibinfo{author}{Raviv, E.}, \bibinfo{author}{Bouwman, K.~E.}, \&
  \bibinfo{author}{van Dijk, D.} (\bibinfo{year}{2015}).
\newblock \bibinfo{title}{Forecasting day-ahead electricity prices: Utilizing
  hourly prices}.
\newblock {\it \bibinfo{journal}{Energy Economics}\/},  {\it
  \bibinfo{volume}{50}\/}, \bibinfo{pages}{227--239}.
\bibitem[{Shafer \& Vovk(2008)}]{shafer2008tutorial}
\bibinfo{author}{Shafer, G.}, \& \bibinfo{author}{Vovk, V.}
  (\bibinfo{year}{2008}).
\newblock \bibinfo{title}{A tutorial on conformal prediction}.
\newblock {\it \bibinfo{journal}{Journal of Machine Learning Research}\/},
  {\it \bibinfo{volume}{9}\/}, \bibinfo{pages}{371--421}.
\bibitem[{Smola \& Sch{\"o}lkopf(2004)}]{smola2004tutorial}
\bibinfo{author}{Smola, A.~J.}, \& \bibinfo{author}{Sch{\"o}lkopf, B.}
  (\bibinfo{year}{2004}).
\newblock \bibinfo{title}{A tutorial on support vector regression}.
\newblock {\it \bibinfo{journal}{Statistics and computing}\/},  {\it
  \bibinfo{volume}{14}\/}, \bibinfo{pages}{199--222}.
\bibitem[{Tibshirani(1996)}]{tibshirani1996regression}
\bibinfo{author}{Tibshirani, R.} (\bibinfo{year}{1996}).
\newblock \bibinfo{title}{Regression shrinkage and selection via the lasso}.
\newblock {\it \bibinfo{journal}{Journal of the Royal Statistical Society.
  Series B (Methodological)}\/},  {\it \bibinfo{volume}{58}\/},
  \bibinfo{pages}{267--288}.
\bibitem[{Uniejewski \& Weron(2018)}]{uniejewski2018efficient}
\bibinfo{author}{Uniejewski, B.}, \& \bibinfo{author}{Weron, R.}
  (\bibinfo{year}{2018}).
\newblock \bibinfo{title}{Efficient forecasting of electricity spot prices with
  expert and lasso models}.
\newblock {\it \bibinfo{journal}{Energies}\/},  {\it \bibinfo{volume}{11}\/},
  \bibinfo{pages}{2039}.
\bibitem[{Uniejewski et~al.(2018)Uniejewski, Weron \&
  Ziel}]{uniejewski2018variance}
\bibinfo{author}{Uniejewski, B.}, \bibinfo{author}{Weron, R.}, \&
  \bibinfo{author}{Ziel, F.} (\bibinfo{year}{2018}).
\newblock \bibinfo{title}{Variance stabilizing transformations for electricity
  spot price forecasting}.
\newblock {\it \bibinfo{journal}{IEEE Transactions on Power Systems}\/},  {\it
  \bibinfo{volume}{33}\/}, \bibinfo{pages}{2219--2229}.
\bibitem[{Vapnik et~al.(1997)Vapnik, Golowich \& Smola}]{vapnik1997support}
\bibinfo{author}{Vapnik, V.}, \bibinfo{author}{Golowich, S.~E.}, \&
  \bibinfo{author}{Smola, A.~J.} (\bibinfo{year}{1997}).
\newblock \bibinfo{title}{Support vector method for function approximation,
  regression estimation and signal processing}.
\newblock In {\it \bibinfo{booktitle}{Advances in neural information processing
  systems}\/} (pp. \bibinfo{pages}{281--287}).
\bibitem[{Vovk(2015)}]{vovk2015cross}
\bibinfo{author}{Vovk, V.} (\bibinfo{year}{2015}).
\newblock \bibinfo{title}{Cross-conformal predictors}.
\newblock {\it \bibinfo{journal}{Annals of Mathematics and Artificial
  Intelligence}\/},  {\it \bibinfo{volume}{74}\/}, \bibinfo{pages}{9--28}.
\bibitem[{Vovk et~al.(2005)Vovk, Gammerman \& Shafer}]{vovk2005algorithmic}
\bibinfo{author}{Vovk, V.}, \bibinfo{author}{Gammerman, A.}, \&
  \bibinfo{author}{Shafer, G.} (\bibinfo{year}{2005}).
\newblock {\it \bibinfo{title}{Algorithmic learning in a random world}\/}.
\newblock \bibinfo{publisher}{Springer Science \& Business Media}.
\bibitem[{Weber(2010)}]{weber2010adequate}
\bibinfo{author}{Weber, C.} (\bibinfo{year}{2010}).
\newblock \bibinfo{title}{Adequate intraday market design to enable the
  integration of wind energy into the european power systems}.
\newblock {\it \bibinfo{journal}{Energy Policy}\/},  {\it
  \bibinfo{volume}{38}\/}, \bibinfo{pages}{3155--3163}.
\bibitem[{Weron(2007)}]{weron2007modeling}
\bibinfo{author}{Weron, R.} (\bibinfo{year}{2007}).
\newblock {\it \bibinfo{title}{Modeling and forecasting electricity loads and
  prices: A statistical approach}\/} volume \bibinfo{volume}{403}.
\newblock \bibinfo{publisher}{John Wiley \& Sons}.
\bibitem[{Weron(2014)}]{weron2014electricity}
\bibinfo{author}{Weron, R.} (\bibinfo{year}{2014}).
\newblock \bibinfo{title}{Electricity price forecasting: A review of the
  state-of-the-art with a look into the future}.
\newblock {\it \bibinfo{journal}{International journal of forecasting}\/},
  {\it \bibinfo{volume}{30}\/}, \bibinfo{pages}{1030--1081}.
\bibitem[{Weron \& Misiorek(2008)}]{weron2008forecasting}
\bibinfo{author}{Weron, R.}, \& \bibinfo{author}{Misiorek, A.}
  (\bibinfo{year}{2008}).
\newblock \bibinfo{title}{Forecasting spot electricity prices: A comparison of
  parametric and semiparametric time series models}.
\newblock {\it \bibinfo{journal}{International journal of forecasting}\/},
  {\it \bibinfo{volume}{24}\/}, \bibinfo{pages}{744--763}.
\bibitem[{Winkler(1972)}]{winkler1972decision}
\bibinfo{author}{Winkler, R.~L.} (\bibinfo{year}{1972}).
\newblock \bibinfo{title}{A decision-theoretic approach to interval
  estimation}.
\newblock {\it \bibinfo{journal}{Journal of the American Statistical
  Association}\/},  {\it \bibinfo{volume}{67}\/}, \bibinfo{pages}{187--191}.
\bibitem[{Wu et~al.(2013)Wu, Chan, Tsui \& Hou}]{wu2013new}
\bibinfo{author}{Wu, H.}, \bibinfo{author}{Chan, S.}, \bibinfo{author}{Tsui,
  K.}, \& \bibinfo{author}{Hou, Y.} (\bibinfo{year}{2013}).
\newblock \bibinfo{title}{A new recursive dynamic factor analysis for point and
  interval forecast of electricity price}.
\newblock {\it \bibinfo{journal}{IEEE Transactions on Power Systems}\/},  {\it
  \bibinfo{volume}{28}\/}, \bibinfo{pages}{2352--2365}.
\bibitem[{Yeo \& Johnson(2000)}]{yeo2000new}
\bibinfo{author}{Yeo, I.-K.}, \& \bibinfo{author}{Johnson, R.~A.}
  (\bibinfo{year}{2000}).
\newblock \bibinfo{title}{A new family of power transformations to improve
  normality or symmetry}.
\newblock {\it \bibinfo{journal}{Biometrika}\/},  {\it \bibinfo{volume}{87}\/},
  \bibinfo{pages}{954--959}.
\bibitem[{Zhang et~al.(2012)Zhang, Tan \& Yang}]{zhang2012day}
\bibinfo{author}{Zhang, J.}, \bibinfo{author}{Tan, Z.}, \&
  \bibinfo{author}{Yang, S.} (\bibinfo{year}{2012}).
\newblock \bibinfo{title}{Day-ahead electricity price forecasting by a new
  hybrid method}.
\newblock {\it \bibinfo{journal}{Computers \& Industrial Engineering}\/},  {\it
  \bibinfo{volume}{63}\/}, \bibinfo{pages}{695--701}.
\bibitem[{Ziel \& Weron(2018)}]{ziel2018day}
\bibinfo{author}{Ziel, F.}, \& \bibinfo{author}{Weron, R.}
  (\bibinfo{year}{2018}).
\newblock \bibinfo{title}{Day-ahead electricity price forecasting with
  high-dimensional structures: Univariate vs. multivariate modeling
  frameworks}.
\newblock {\it \bibinfo{journal}{Energy Economics}\/},  {\it
  \bibinfo{volume}{70}\/}, \bibinfo{pages}{396--420}.

\end{thebibliography}

\endgroup
\end{document}